\documentclass[submission,copyright,creativecommons]{eptcs}
\providecommand{\event}{TYPES 2009 Post-Proceedings} % Name of the event you are submitting to
\usepackage{breakurl}             % Not needed if you use pdflatex only.

\include{amssymb}
\newcommand{\selfcomment}[1]{\ifodd 0 {\sf #1 }\fi}
\newcommand{\selfc}{\selfcomment}
\newcommand{\selfcc}[1]{\ifodd 1 {\sf #1 }\fi}
\newcommand{\selffootnote}[1]{\ifodd 0 \footnote{{\sf #1}} \fi}

\newtheorem{definition}{Definition}[section]
\newtheorem{defn}[definition]{Definition}
\newtheorem{theorem}[definition]{Theorem}
\newtheorem{thm}[definition]{Theorem}
\newtheorem{lemma}[definition]{Lemma}
\newtheorem{cor}[definition]{Corollary}
\newtheorem{corollary}[definition]{Corollary}

\newcommand{\remark}{\ \\{\bf Remark}\ }

\newcommand{\qed}{}%\quad\hfill\mbox{$\Box$}}

\mathcode`:="003A               % ordinary colon (tight spacing)
\mathchardef\colon="303A        % relation colon, rule colon
\newcommand{\G}{\Gamma}
\newcommand{\ts}{\vdash}        % turnstile
\newcommand{\lam}{\lambda}
\newcommand{\To}{\Rightarrow}                   % imply -- double right arrow
\newcommand{\IH}{induction hypothesis}  % IH for induction hypothesis
\newcommand{\ML}{Martin-L\"{o}f}        % Martin-Lof
\newcommand{\SN}{strongly normalisable} % SN abbr. strongly normalizable
\newcommand{\eg}{{e.g.}}

\newcommand{\ie}{{i.e.}}

\newcommand{\wrt}{{w.r.t.}}

\newcommand{\Frule}[4]
{\[ \hbox to \columnwidth
    { \rlap{$#4$} \hfil $
      {{\displaystyle\strut #1}\over{\displaystyle\strut #2}}
      \quad\makebox[0pt][l]{\it #3} $
      \hfil }
\]}
\newcommand{\tworule}[4]                % special for LF-rules picture
{\[ {{#1}\over{#2}}
    \ \
    {{#3}\over{#4}}
\]}
\newcommand{\threerule}[6]              % special for LF-rules picture
{\[ {{#1}\over{#2}}
    \ \ \
    {{#3}\over{#4}}
    \ \ \
    {{#5}\over{#6}}
\]}
\renewcommand{\SN}{strongly normalisable}

\newcommand{\record}[1]{\langle #1 \rangle}

\title{Typed Operational Semantics for Dependent Record Types\thanks{The authors acknowledge the support from the Leverhulme Trust (grant ref. F/07-537/AA) and the first author also acknowledges the support from the College Research Studentship and an award from Computer Science Department in Royal Holloway, University of London.}}
\author{Yangyue Feng and Zhaohui Luo
\institute{Department of Computer Science\\ Royal Holloway, University of London \\
Egham, Surrey TW20 0EX, UK\\}
\email{\{yangyue,zhaohui\}@cs.rhul.ac.uk}
}
\def\titlerunning{Typed Operational Semantics for Dependent Record Types}
\def\authorrunning{Yangyue Feng and Zhaohui Luo}

\begin{document}

\maketitle

\begin{abstract}
Typed operational semantics is a method developed by H. Goguen to prove meta-theoretic properties of type systems. This paper studies the metatheory of a type system with dependent record types, using the approach of typed operational semantics. In particular, the metatheoretical properties we have proved include strong normalisation, Church-Rosser and subject reduction.
\end{abstract}

\section{Introduction}

\par{H. Goguen \cite{healf:thesis,healf:TLCA99YY} has developed a method called \emph{typed operational semantics} (TOS for short) to prove meta-theoretic properties of type theories, including strong normalisation, Church-Rosser and subject reduction. In this paper, using the TOS approach, we study the meta-theoretic properties of a type system with dependent record types. }

\par{A record type is a type of labelled tuples called records.  A dependent record type (DRT) is a type of records whose fields may have types that depend on the values of earlier fields.  Dependent records have been studied previously for various different type systems \cite{Harper-Lillibridge93, bet-tar:subtyping98, Pollack:records02, ctp:semantic-records05}, with applications to the study of module mechanisms for both programming and proof languages. Recently, in the context of studying manifest fields of module types, the second author has proposed a formulation of dependent record types \cite{luo:TYPES08}, for type theories with canonical objects such as \ML's type theory, and shown in \cite{luo:MLPA09} that, in some applications, dependent record types are more useful than $\Sigma$-types (dependent types of tuples without labels). }

Studying the meta-theory of dependent record types, the contributions of the current paper are two-fold.  First of all, the meta-theory of dependent record types has not been well-studied.  This work makes a positive contribution, showing that our formulation of dependent record types has the good meta-theoretic properties such as strong normalisation.  Secondly, the type theory we study has record \emph{types} as studied in \cite{Pollack:records02,luo:TYPES08}, rather than record \emph{kinds} as in \cite{bet-tar:subtyping98,ctp:semantic-records05}.  Since types have a much more sophisticated structure than kinds, the meta-theory for dependent record types is expected to be much more difficult than that for dependent record kinds as found in, \eg, \cite{ctp:semantic-records05}.  We shall study the meta-theory by taking the TOS approach, which is shown to be robust enough to deal with dependent record types.  In particular, we study the \emph{intensional} DRTs, that is, the dependent record types without the so-called weakly extensional rules (these rules are considered in \cite{luo:TYPES08}).  The typed operational semantics for intensional DRTs is developed and shown to be sound and complete and, based on this, it is proved that the intensional DRTs have good meta-theoretic properties, including strong normalisation, Church-Rosser and subject reduction.

The paper is arranged as follows. The type system IDRT for intensional DRTs is described in Section~\ref{sec:LFwithDRT}.  In Section~\ref{sec:TOS}, after introducing the basic idea of TOS, we define the TOS for dependent record types.  The properties of the TOS are studied in Section~\ref{sec:metaTOS} and the meta-theoretic properties of IDRT in Section~\ref{sec:metaDRT}.  Discussions of related work and future work are given in the conclusion.

\section{Dependent Record Types}
\label{sec:LFwithDRT}

A dependent record type is a type of labelled tuples whose fields may have types that depend on the values of earlier fields. For instance, if $Nat$ and $Vect(n)$ are the types of natural numbers and vectors of length $n$, respectively, $\record{n\colon Nat,\ v\colon Vect(n)}$ is the dependent record type with objects (called {\sl records}) such as $\record{n=2,\ v=[5,6]}$, where dependency is respected: the vector $[5,6]$ must be of type $Vect(2)$.

Formally, in our study, dependent record types are formulated as an extension of the logical framework that we describe briefly first.

\paragraph{Logical Framework.}  LF \cite{luo:book94} is the typed version of Martin-L\"{o}f's logical framework \cite{NPS:book}.  It is itself a type system that serves as a meta-language to specify type theories such as \ML's intensional type theory \cite{NPS:book} and the Unifying Theory of dependent Types (UTT) \cite{luo:book94}.  Here, we give only a brief introduction, fixing the notations to be used in the paper.  (For details of, for example, how inductive types like $Nat$, $\Pi$-types and $\Sigma$-types can be specified in the logical framework, see Part III of \cite{NPS:book} or Chapter 9 of \cite{luo:book94}.)

In LF, the syntactical entities \emph{contexts}, \emph{kinds} and \emph{terms} are of the following forms:
\begin{eqnarray*}
Contexts \ & \ \Gamma & ::= \ \ \ () \ \ | \ \ \Gamma, x \colon A \\
LF\ Kinds \ & \ K & ::= \ \ \ Type \ \ | \ \ El(A) \ \ | \ \ (x:K)K' \\ %\ | \
LF\ Terms \ & \ M & ::= \ \ \ x \ \ | \ \ [x:K]M \ \ | \ \ M(M')
\end{eqnarray*}
The types in LF are called \emph{kinds}, including:
\begin{itemize}
\item $Type$ -- the kind representing the collection of all types ($A$ is a type if $A \colon Type$);
\item $El(A)$ -- the kind of objects of type $A$ (we often omit $El$); and
\item $(x:K)K'$ (or simply $(K)K'$ when $x \notin FV (K')$) -- the kind of dependent functional operations.
\end{itemize}
The judgement forms in LF include, for example,
\begin{itemize}
  \item $\G\ts k\colon K$, which asserts that $k$ is an object of kind $K$; and
  \item $\G\ts k=k'\colon K$, which asserts that $k$ and $k'$ are (computationally) equal objects of kind $K$.
\end{itemize}
The inference rules of LF to define the typing relation and the computational equality are given in Appendix~\ref{app:LF-rules}.  In particular, $\beta\eta$-equal objects are computationally equal.  For instance, an abstraction $[x:K]M$ can be applied to form $([x:K]M)(a)$ that is computationally equal to $[a/x]M$.

\begin{notation}
We shall use $\equiv$ to denote the syntactical identity (up to $\alpha$-conversion).
\end{notation}

\paragraph{Dependent Record Types.}
We now give a formal presentation of the system IDRT of intensional dependent record types, which is an extension of LF.  The syntax of this type system is given as follows, where $\mathcal{L}$ is an (infinite) set of labels, $l\in \mathcal{L}$ and $L\subset\mathcal{L}$ is finite:
\begin{eqnarray*}
%Contexts \ & \ C & ::= \ \ \ () \ | \ \Gamma, x:A \\
%LF\ Terms \ & \ K & ::= \ \ \ Type \ | \ \textbf{El}(M) \ | \ (x:K)K' \ | \ [x:K]K' \ | \ f(k) \\
%LF\ Types \ & \ T & ::= \ \ \ x \ | \ [x:K]K' \ | \ f(k)  \\
Kinds\ of\ Record\ Types \ & K_R & ::= \ \ \ RType \ | \ RType[L] \\ %\ (L \subseteq \mathcal{L}) \\
Record\ Types \ & \ R & ::= \ \ \ \langle \rangle \ | \ \langle R, \ l \colon A \rangle  \\
Records \ & \ r & ::= \ \ \ \langle \rangle \ | \ \langle r,\ l=a \colon A \rangle %\\
\end{eqnarray*}
The inference rules of IDRT consist of the rules for LF (Appendix~\ref{app:LF-rules}) and the additional rules in Figure~\ref{DRT-rules}. Here are some informal explanations.
\begin{figure}[top]
\framebox[5.8in][l]{
\begin{minipage}{\linewidth}
\ \\

\ \ \ \emph{Kinds of record types}
$$
\frac{\Gamma \: valid}{\Gamma \vdash RType \: kind}\: \: \: \:
\frac{\Gamma \: valid \: \: \: %L \subseteq \mathcal{L}
}{\Gamma \vdash RType[L] \: kind}
$$
$$
\frac{\Gamma \vdash R \colon RType[L] \: \: \: L \subseteq L' %\subseteq \mathcal{L}
}{\Gamma \vdash R \colon RType[L']} \: \: \: \:
\frac{\Gamma \vdash R \colon RType[L]}{\Gamma \vdash R \colon RType} \: \: \: \:
\frac{\Gamma \vdash R \colon RType}{\Gamma \vdash R \colon Type}
$$

\ \ \ \emph{Formation rules}
$$
\frac{\Gamma \: valid}{\Gamma \vdash \langle \rangle \colon RType[\emptyset]}
\: \: \: \:
\frac{\Gamma \vdash R \colon RType[L] \: \: \: \Gamma \vdash A \colon (R)Type \: \: \: l \notin L}{\Gamma \vdash \langle R, \ l \colon A \rangle \colon RType[L \cup \{l\}]}
$$

\ \ \ \emph{Introduction rules}
$$
\frac{\Gamma \: valid}{\Gamma \vdash \langle \rangle \colon \langle \rangle}
\: \: \: \:
\frac{\Gamma \vdash \langle R, \ l\colon A \rangle \colon RType \: \: \: \Gamma \vdash r \colon R \: \: \: \Gamma \vdash a \colon A(r)}{\Gamma \vdash \langle r, \ l=a \colon A \rangle \colon \langle R, \ l \colon A \rangle}
$$

\ \ \ \emph{Elimination rules}
$$
\frac{\Gamma \vdash r \colon \langle R, \ l \colon A \rangle}{\Gamma \vdash [r]\colon R}
\: \: \: \:
\frac{\Gamma \vdash r \colon \langle R, \ l \colon A \rangle}{\Gamma \vdash r.l \colon A([r])}
$$
$$
\frac{\Gamma \vdash r \colon \langle R, \ l \colon A \rangle \: \: \: \Gamma \vdash [r].l' \colon B \: \: \: l \neq l'}{\Gamma \vdash r.l' \colon B}
$$

\ \ \ \emph{Computation rules}
$$
\frac{\Gamma \vdash \langle r, \ l = a \colon A \rangle \colon \langle R, \ l \colon A \rangle}{\Gamma \vdash [\langle r, \ l = a \colon A \rangle] = r \colon R}
\: \: \: \:
\frac{\Gamma \vdash \langle r, \ l = a \colon A \rangle \colon \langle R, \ l \colon A \rangle}{\Gamma \vdash \langle r, \ l = a \colon A \rangle . l = a \colon A(r)}
$$
$$
\frac{\Gamma \vdash r \colon \langle R, \ l \colon A \rangle \: \: \: \Gamma \vdash [r].l' \colon B \: \: \: l \neq l'}{\Gamma \vdash r.l' = [r].l' \colon B}
$$
%%%%%%%%% May 10th remove %%%%%%%%%
%\ \ \ \emph{Weakly extensional equality rules}
%$$
%\frac{\Gamma \vdash r\colon \langle \rangle}{\Gamma \vdash r = \langle \rangle \colon \langle \rangle}
%\: \: \: \:
%\frac{\begin{array}{c} \Gamma \vdash r \colon \langle R, \ l\colon A \rangle \: \: \: \Gamma \vdash r' \colon \langle R, \ l\colon A \rangle \\ \Gamma \vdash [r] = [r']\colon R \: \: \: \Gamma \vdash r.l = r'.l \colon A([r]) \end{array}}{\Gamma \vdash r = r' \colon \langle R, \ l\colon A \rangle}
%$$

\ \ \ \emph{Congruence rules for record types}
$$
\frac{\Gamma \: valid}{\Gamma \vdash \langle \rangle = \langle \rangle \colon RType[\emptyset]}
\: \: \: \:
\frac{\Gamma \vdash R = R' \colon RType[L] \: \: \: \Gamma \vdash A = A' \colon (R)Type \; \: \: l \notin L}{\Gamma \vdash \langle R, \ l \colon A \rangle = \langle R', \ l \colon A' \rangle \colon RType[L \cup \{l\}]}
$$

\ \ \ \emph{Congruence rules for records}
$$
\frac{\Gamma \: valid}{\Gamma \vdash \langle \rangle = \langle \rangle \colon \langle \rangle}
\: \: \: \:
\frac{\begin{array}{c} \Gamma \vdash R \colon RType[L] \: \: \: l \notin L \\ \Gamma \vdash r = r' \colon R \: \: \: \Gamma \vdash a = a' \colon A(r) \: \: \: \Gamma \vdash A = A' \colon (R)Type \end{array}}{\Gamma \vdash \langle r, \ l = a \colon A \rangle = \langle r', \ l = a' \colon A' \rangle \colon \langle R, \ l \colon A \rangle}
$$
$$
\frac{\Gamma \vdash r = r' \colon \langle R, l \colon A \rangle}{\Gamma \vdash [r] = [r'] \colon R}
\: \: \: \:
\frac{\Gamma \vdash r = r' \colon \langle R, l \colon A \rangle}{\Gamma \vdash r.l = r'.l \colon A([r])}
$$
\\
\end{minipage}
} %framebox
\caption{Inference Rules of IDRT} \label{DRT-rules}
\end{figure}

\begin{itemize}
  \item We add new kinds $RType$ and $RType[L]$ of record types.  Intuitively, $RType[L]$ is the kind of the record types whose (top-level) labels are all in $L$, a finite set of labels.  Naturally, if $L\subseteq L'$, every record type in $RType[L]$ is also in $RType[L']$.  The kind $RType$ is the kind of all record types and could  conceptually be understood as `$RType[\mathcal{L}]$'.  Finally, every record type is also a type.  These are formally reflected in the rules for the kinds of record types in Figure~\ref{DRT-rules}.
  \item Record types are types of the form $\record{}$ or $\record{R,\ l\colon A}$.  Intuitively, a record type is of the form $\record{l_1\colon A_1,\ ...,\ l_n\colon A_n}$,\footnote{We overload the $\record{\ ...\ }$ notation for records and their types.  It is always possible to distinguish between the two.} where each $l_i\colon A_i$ is a \emph{field} labelled by $l$.  An object of this record type is a labelled tuple $\record{l_1 = a_1\colon A_1,\ ..., \ l_n = a_n\colon A_n}$, where $a_i$ is of the type of the corresponding field.  \selfc{The kind of the field types are dependent kinds from \emph{previous} record types to $Type$.}  Note that, formally, each $A_i$ in the record type is not a type, but a family of types; this is how dependency is incorporated -- we have dependent record types.
      
      Notation-wise, we shall adopt the following notational conventions: for record types, we write $\langle l_1 \colon A_1,\ ...,\ l_n \colon A_n \rangle$ for $\langle \langle \langle \rangle,\ l_1 \colon A_1 \rangle,\ ...,\ l_n \colon A_n \rangle$ and often use label occurrences/non-occurrences to show dependency/non-dependency respectively.
For instance, we write $\langle n \colon Nat, \ v \colon Vect(n) \rangle$ for $\langle \langle \langle \rangle,\ n \colon NAT \rangle, \ v \colon [x:\langle n \colon NAT \rangle] Vect(x.n) \rangle$ where $NAT \equiv [\_\ :\langle \rangle]Nat$, and $\langle R, \ l \colon Vect(2) \rangle$ for $\langle R, \ l \colon [\_\ :R]Vect(2) \rangle$.
  \item There are two operations on records: \emph{restriction} (or first projection) $[r]$ that removes the last component of record $r$ and \emph{field selection} $r.l$ that selects the value of the field labelled by $l$.  For instance, intuitively, for the record $r \equiv \record{l_1 = a_1\colon A_1,\ l_2 = a_2\colon A_2, \ l_3 = a_3\colon A_3}$ of type $\record{l_1\colon A_1,\ l_2\colon A_2,\ l_3\colon A_3}$, we have $[r] = \record{l_1\colon A_1,\ l_2\colon A_2}$ and $r.l_2 = [r].l_2 = a_2$.  These are formally reflected in the introduction, elimination and computation rules in Figure~\ref{DRT-rules}.
  \item The congruence rules for record types and records in Figure~\ref{DRT-rules} propagate the computational equality through the term structure.  Also, we do not include the weakly extensional equality rules as considered in \cite{luo:TYPES08}.  Therefore, we call the system the type system for intensional DRTs.
\end{itemize}
We shall adopt the following terminology: the terms of the form $\record{r,\ l=a\colon A}$ will be called \emph{pair-records}.  (For example, we shall use this terminology in specifying the TOS-rules for record types in Figure~\ref{TOS-DRT-rules} in Section~\ref{sec:TOS-IDRT}.)

\paragraph{Record types v.s. record kinds.}  It is worth pointing out that our type system contains dependent record \emph{types} (as studied by Pollack \cite{Pollack:records02}, Luo \cite{luo:TYPES08,luo:MLPA09} and the current paper), rather than dependent record \emph{kinds} (as studied by Betarte and Tasistro \cite{bet-tar:subtyping98} and Coquand, Pollack and Takeyama \cite{ctp:semantic-records05}\footnote{\emph{Types} in the terminology of \ML's type theory are what we call \emph{kinds} in this paper.  Therefore, the so-called record types in \cite{bet-tar:subtyping98} and \cite{ctp:semantic-records05} are really record kinds.}).  We would like to distinguish these two notions clearly: in a type theory with inductive types, types include those such as $Nat$ of natural numbers and $\Sigma$-types of dependent pairs, while the examples of kinds include, for example, the kind $Type$ of all types.  They exist at two completely different levels and have rather different structures and properties.

In general, types have a much more sophisticated and richer structure than kinds.  For instance, it is easy to show that a kind is of the form either $Type$ or $(x:K)K'$, but types are not (\eg, a type may be of the form $f(a)$).  To appreciate the difference, let us consider the issue of ensuring label distinctness.  If one considers only record kinds, it is easy to guarantee that the labels in the same record kind are distinct because of the limited syntactic forms of kinds (see, for example, \cite{ctp:semantic-records05}).  However, this is not easy at all for record types (think, for example, how one ensures that a label does not occur in a type of the form $f(a)$).  In our case, we have to introduce the kinds $RType[L]$ to ensure that it is the case that the (top-level) labels in the same record type are distinct.  In other words, intuitively, $l\not= l'$ for any record type $\record{...,\ l\colon A,\ ...,\ l'\colon A',\ ...}$.  This is guaranteed by means of the side condition $\l\not\in L$ of the second formation rule in Figure~\ref{DRT-rules}.

That a type system with record types is more powerful than one with only record kinds can be understood from another angle when one wants to introduce universes of record types.  It is possible to introduce type universes for dependent record types, as shown in \cite{luo:MLPA09}; this, however, cannot be done for record kinds.  Therefore, record types are more useful than record kinds (for example, in representing module types in data refinement \cite{luo:MLPA09}).

Since types have a more sophisticated structure than kinds, it is more difficult to study the meta-theoretic properties of a system with record types, as compared with a meta-theoretic study of record kinds.  As we show in this paper, the approach of using typed operational semantics can be used in this endeavour.

\section{Typed Operational Semantics for Dependent Record Types}
\label{sec:TOS}

The typed operational semantics (TOS for short) is a proof-theoretic method to prove the meta-theoretic properties of type theories.  It was developed by H. Goguen in his PhD thesis \cite{healf:thesis}, where he studied the meta-theory of UTT and proved that UTT has the nice properties such as Church-Rosser, Subject Reduction and Strong Normalisation.

In this paper, the TOS approach is applied to study the meta-theory of dependent record types.  After a brief informal introduction of the approach, we develop the typed operational semantics for the system IDRT of intensional DRTs and show that it has the soundness and completeness properties.  The meta-theoretic properties of dependent record types are studied in the next section.

\subsection{The TOS Approach}
\label{sec:introTOS}

For a type theory, its typed operational semantics captures its computational behaviour, usually given by its (untyped) reduction relation.  For example, in TOS, the following judgement
\[ \G\models M\to N\to P\colon A \]
informally asserts that, among other things, $N$ and $P$ are the weak-head normal form and the normal form of the term $M$, respectively.\footnote{Formally, the reduction relation and the TOS are related to each other by means of the `adequacy theorems' such as Lemmas~\ref{AUR} and~\ref{ANF} for IDRT in Section~\ref{subsec: adeq}.}  For the logical framework LF, for example, its corresponding TOS has been studied \cite{healf:TLCA99YY} and its inference rules are given in Appendix~\ref{app:LF-TOSrules}.  Since many meta-theoretic properties of a type theory are concerned with its computational behaviour, it is not a surprise that TOS provides an effective approach to the meta-theory of type theories.\footnote{It is worth noting that, although it is useful to study the meta-theory for many type theories, the TOS approach would not be suitable for non-normalising type theories.  See \cite{healf:thesis} for discussions.}

The TOS and its corresponding type theory are related to each other by means of the soundness and completeness theorems.  Using the judgement $\G\models M\colon A$ to abbreviate `$\G\models M\to N\to P\colon A$ for some $N$ and $P$', we can state the soundness and completeness properties as follows:
\begin{itemize}
  \item Soundness: $\G\ts M\colon A$ implies $\G\models M\colon A'$ (for $A'$ that is the `normal form' of $A$).
  \item Completeness: $\G\models M\colon A$ implies $\G\ts M\colon A$.
\end{itemize}
Based on soundness and completeness, we can prove many meta-theoretic properties of the type theory.  For example, it can be shown that, if $\G\models M\colon A'$, then $M$ is strongly normalisable.  Therefore, strong normalisation, the property that every well-typed term is strongly normalisable, can be proved by means of such a fact together with the soundness property, as pictured as follows:

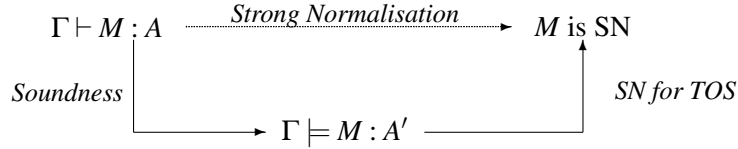
\begin{picture}(100, 80)(-60, -50)
% \put(5, 1){an armadillo}
\put(30, 10){\makebox(0,0){$\Gamma \ts M \colon A$}}
\put(120, -30){\makebox(0,0){$\Gamma \models M \colon A'$}}
\put(210, 10){\makebox(0,0){$M$ is SN}}
\put(40, -30){\line(0, 1){35}}
\put(40, -30){\vector(1, 0){50}}
\put(150, -30){\line(1, 0){60}}
\put(210, -30){\vector(0, 1){35}}
\put(60, 10){\dashbox{.5}(120, 0)[t]{}}%{\vector(1, 0){120}}
\put(180, 10){\vector(1, 0){3}}
\put(120, 14){\makebox(0,0){\emph{\begin{small}Strong Normalisation\end{small}}}}
\put(15, -15){\makebox(0,0){\emph{\begin{small}Soundness\end{small}}}}
%\put(252, -10){\makebox(0,0){\emph{\begin{small}Adequacy w.r.t \end{small}}}}
\put(245, -15){\makebox(0,0){\emph{\begin{small}SN for TOS\end{small}}}}
\end{picture}

\noindent As shown in this paper, for dependent record types, the SN property for the corresponding TOS is proven in Theorem~\ref{SN-TOS}.  Then, by the Soundness Theorem (Theorem~\ref{Soundness-TOS-LF}), we can show that strong normalisation for IDRT (Corollary~\ref{SN-LFDRT}).

Note that, to implement such ideas is not a simple matter: it requires one to prove:
\begin{itemize}
  \item that the TOS is `adequate' \wrt\ the (untyped) reduction relation,
  \item that the TOS is sound and complete \wrt\ the original type theory, and
  \item that the TOS satisfies some specific meta-theoretic properties (\eg, strong normalisation).
\end{itemize}
Then, one can transfer the results to the original type theory to show that it has nice meta-theoretic properties.  This is what we shall do for IDRT, the type theory with dependent record types.

\subsection{TOS for Dependent Record Types}
\label{sec:TOS-IDRT}

The typed operational semantics for dependent record types is described in this section.  The judgement forms in a TOS are given in Figure~\ref{TOS-judgements},
\begin{figure}[top]
%\framebox[5.2in][l]{
\begin{minipage}{\linewidth}
\begin{eqnarray*}
Basic\ forms: \ \ \ & \ \ \ \ \ \ \ \ Abbreviated\ forms: \\
\models \Gamma \rightarrow \Delta \ \ \ & \ \ \ \ \ \ \ \ \Gamma \models\ ok \\
%& Context\ Normalizing\ \\
\Gamma \models A \rightarrow B \ \ \ & \ \ \ \ \ \ \ \ \Gamma \models M \rightarrow_{w} N \colon A \\
%& Kind\ Normalizing\ \\
\Gamma \models M \rightarrow N \rightarrow P \colon A \ \ \ & \ \ \ \ \ \ \ \ \Gamma \models M \rightarrow_{n} P \colon A \\
%& Term\ Weak-Head\ Reducing\ and\ Normalizing\ \\
\ \ \ & \ \ \ \ \ \ \ \ \Gamma \models M \colon A
%\Gamma \models ok
%& Context\ Normalizable \ \\
%\Gamma \models M \rightarrow_{w} N \colon A
%& Term\ Weak-Head\ Reducible \ \\
%\Gamma \models M \rightarrow_{n} P \colon A
%& Term\ Normalizable\ \ \\
%\Gamma \models M \colon A
%& Term\ Well-typed\ and\ Normalizable\ \\
\end{eqnarray*}
\end{minipage}
%}
\caption{Judgement Forms in Typed Operational Semantics} \label{TOS-judgements}
\end{figure}
three of which are the basic forms of judgements whose informal meanings are:
\begin{itemize}
  \item $\models \Gamma \rightarrow \Delta$: the context $\Gamma$ has context $\Delta$ as its normal form;
  \item $\Gamma \models A \rightarrow B$: the kind $A$ is well-formed in context $\Gamma$ and has normal form $B$; and
  \item $\Gamma \models M \rightarrow N \rightarrow P \colon A$: the terms $M$, $N$, $P$ are well-formed in context $\Gamma$ of kind $A$ and $M$ has weak-head normal form $N$ and normal form $P$.
\end{itemize}
From these basic judgements, one can define other forms of judgements, including the following:
\begin{itemize}
  \item $\Gamma \models\ ok$ stands for `$\models \Gamma \rightarrow \Delta$ for some $\Delta$';
  \item $\Gamma \models M \rightarrow_{w} N \colon A$ stands for `$\Gamma \models M \rightarrow N \rightarrow P \colon A$ for some $P$';
  \item $\Gamma \models M \rightarrow_{n} P \colon A$ stands for `$\Gamma \models M \rightarrow N \rightarrow P \colon A$ for some $N$'; and
  \item $\Gamma \models M \colon A$ stands for `$\Gamma \models M \rightarrow N \rightarrow P \colon A$ for some  $N$ and $P$'.
\end{itemize}

The typed operational semantics for the type system IDRT of intensional DRTs is the extension of that for LF (Appendix~\ref{app:LF-TOSrules}) with the inference rules given in Figure~\ref{TOS-DRT-rules}.
\begin{figure}[top]
\framebox[6.2in][l]{
\begin{minipage}{\linewidth}
\ \\

\ \ \ \emph{Record Kinds}
$$
\frac{\Gamma \models ok}{\Gamma \models RType \rightarrow RType} \ RTYPE \: \: \: \: \: \:
\frac{\Gamma \models ok \: \: \: %L \subseteq \mathcal{L}
}{\Gamma \models RType[L] \rightarrow RType[L]} \ RTYPE[L]
$$
\ \ \ \emph{Record Types}
$$
\frac{\Gamma \models ok}{\Gamma \models \langle \rangle \rightarrow \langle \rangle \rightarrow \langle \rangle \colon RType[\emptyset]} \ \ EMP_{RCDT} %\: \: \: \: \: \:
$$
$$
\frac{\Gamma \models R \rightarrow_n P \colon RType[L] \: \: \: \: \Gamma \models A \rightarrow_n B \colon (P)Type \: \: \: \: l \notin L}{\Gamma \models \langle R, l \colon A \rangle \rightarrow \langle R, l \colon A \rangle \rightarrow \langle P, l \colon B \rangle \colon RType[L \cup \{l\}]} \ \ RCDT
$$
\ \ \ \emph{Pair-records}
$$
\frac{\Gamma \models ok}{\Gamma \models \langle \rangle \rightarrow \langle \rangle \rightarrow \langle \rangle \colon \langle \rangle} \ \ EMP_{RCD}
$$
$$
\frac{\begin{array}{c} \Gamma \models \langle R, l \colon A \rangle \rightarrow_n \langle P, l \colon B \rangle \colon RType \: \: \: \: \Gamma \models r \rightarrow_n p \colon P \\ \Gamma \models A(r) \rightarrow_n C \colon Type \: \: \: \Gamma \models a \rightarrow_n b \colon C \: \: \: \:
\end{array}}{\Gamma \models \langle r, l=a \colon A \rangle \rightarrow \langle r, l=a \colon A \rangle \rightarrow \langle p, l=b \colon B \rangle \colon \langle P, l \colon B \rangle} \ RCD
$$
\ \ \ \emph{Restrictions}
$$
\frac{\Gamma \models r \rightarrow q \rightarrow p \colon \langle P, l \colon B \rangle \: \: \: \: p,\ q\ not\ pair\texttt{-}records}{\Gamma \models [r] \rightarrow [q] \rightarrow [p] \colon P} \ \ BASE_{RESTR}
$$
$$
\frac{\Gamma \models r \rightarrow_w \langle p, l = b \colon A \rangle \colon \langle P, l \colon B \rangle \: \: \: \: \Gamma \models p \rightarrow s \rightarrow t \colon P \selfc{\: \: \: \: \Gamma \models P \rightarrow_n C \colon Q}}{\Gamma \models [r] \rightarrow s \rightarrow t \colon P} \ RESTR
$$
\ \ \ \emph{Selections}
$$
\frac{\begin{array}{c} \Gamma \models r \rightarrow q \rightarrow p \colon \langle P, l \colon B \rangle \: \: \: \: p,\ q\ not\ pair\texttt{-}records \\ \Gamma \models B([r]) \rightarrow_n C \colon Type \end{array}}{\Gamma \models r.l \rightarrow q.l \rightarrow p.l \colon C} \ BASE_{FLDSEL}
$$
  $$
    \frac{\Gamma \models r \rightarrow_w  \langle p,\ l = b \colon A \rangle \colon \langle P,\ l \colon B \rangle \: \: \: \: \Gamma \models b \rightarrow c \rightarrow d \colon C \: \: \: \: \Gamma \models A(p) \rightarrow_n C \colon Type}{\Gamma \models r.l \rightarrow c \rightarrow d \colon C} \ FLDSEL
  $$
$$
\frac{\begin{array}{c} \Gamma \models r \rightarrow_n s \colon \langle P, l \colon B \rangle \: \: \: \: \Gamma \models [r].l' \rightarrow c \rightarrow d \colon C \: \: \: \: \: l \neq l' \end{array}}{\Gamma \models r.l' \rightarrow c \rightarrow d \colon C}\ FLDSL'
$$
\\
\end{minipage}
}%\framebox
\caption{Inference Rules of Typed Operational Semantics for IDRT} \label{TOS-DRT-rules}
\end{figure}
Most of the rules are self-explanatory. We only mention that, besides using the abbreviated forms of judgement (see above) in the rules, we also use the terminology of `pair-record' as introduced in Section~\ref{sec:LFwithDRT}.  For example, in $(BASE_{RESTR})$, we require that $p$ or $q$ be not a pair-record, for otherwise, for instance, $[p]$ could be a redex and would not be in normal form.

\section{Properties of TOS for Dependent Record Types}
\label{sec:metaTOS}

We shall study the properties of the TOS for IDRT, as presented above in Section~\ref{sec:TOS-IDRT}.  These include those properties \wrt\ the relationship with IDRT (soundness and completeness) and those \wrt\ the reduction relation.
\subsection{Basic Structural Properties}
\label{subsec:struc}

The typed operational semantics satisfy some basic properties as stated in the following lemma, which can all be proved by induction on the TOS-derivations.\footnote{Some of the lemmas (\eg, the strengthening lemma) can only be proved by proving a stronger statement by induction on derivations.  We omit the details here.}

\begin{lemma} \label{Stru-prop}\
\begin{enumerate}
  \item (Context Validity) Any derivation of $\Gamma_0, \Gamma_1 \models J$ has a sub-derivation of $\Gamma_0 \models ok$.
  \item (Variables) Let $dom\{\Gamma\}$ be the set of variables declared in context $\G$ and $FV(M)$ the set of free variables occurring in term $M$.
  \begin{enumerate}
    \item If $\models \Gamma \rightarrow \Delta$, then $dom\{\Delta\} = dom\{\Gamma\}$.
    \item If $\Gamma \models A \rightarrow B$, then $FV(A) \cup FV(B) \subseteq dom\{\Gamma\}$.
    \item If $\Gamma \models M \rightarrow N \rightarrow P \colon A$, then $FV(M) \cup FV(N) \cup FV(P) \cup FV(A)\subseteq dom\{\Gamma\}$.
  \end{enumerate}
\selfc{%I eliminated this, since we have not defined renaming ... (ZL)
  \item (Renaming) Suppose $\gamma$ is a renaming from $\Delta$ to $\Gamma$. If $\Gamma \models J$, then $\Delta \models J[\gamma]$ (where $J[\gamma]$ is the $\gamma$-renaming from $J$, i.e.\ $A[\gamma] \rightarrow B[\gamma]$ or $M[\gamma] \rightarrow N[\gamma] \rightarrow P[\gamma] \colon B[\gamma]$ when $J$ appears as the form $A \rightarrow B$ or $M \rightarrow N \rightarrow P \colon B$ respectively).
}%selfc
  \item (Weakening) If $\Gamma \models J$ and $\Gamma, \Delta\models ok$, then $\G,\Delta \models J$.
  \item (Strengthening) If $\Gamma_0, z:C, \Gamma_1 \models J$ and $z\not\in FV(\G_1)\cup FV(J)$, then $\Gamma_0, \Gamma_1 \models J$.
  \item (Determinacy) \label{UNF}
  \begin{itemize}
    \item If $\models \Gamma \rightarrow \Delta$ and $\models \Gamma \rightarrow \Phi$, then $\Delta \equiv \Phi$.
    \item If $\Gamma \models A \rightarrow B$ and $\Gamma \models A \rightarrow C$, then $B \equiv C$.
    \item If $\Gamma \models M \rightarrow N \rightarrow P \colon B$ and $\Gamma \models M \rightarrow Q \rightarrow R \colon C$, then $N \equiv Q$, $P \equiv R$ and $B \equiv C$.
  \end{itemize}
\end{enumerate}
\end{lemma}

\remark\ The above Lemma~\ref{Stru-prop}(\ref{UNF}) of `Determinacy' says that the TOS-normal forms are unique.  Of course, in order to show that the normal form of a well-typed term (under the usual reduction relation) is unique, one has to prove that the TOS-reductions are adequate.  This is what we do in the following subsection.

\subsection{Adequacy \wrt\ the Untyped Reduction}
\label{subsec: adeq}

We shall show in this section that the notions of computation captured in TOS are adequate \wrt\ the usual (untyped) reduction relation, which is defined in the following definition.

\begin{defn}[Untyped Reduction for IDRT] \label{untyped-red}
The (untyped) one-step reduction over terms, notation $\to$, is the compatible closure\footnote{The compatible closure of a relation $R$ over terms propagates R to all of the terms.  We omit its formal definition here; see \cite{healf:thesis,Fen10} for formal details.} of the relation given by the following rules:
\begin{eqnarray*}
(\beta) \ \ \ \ \ \ \ \ \ \ ([x:A]M)N & \rightarrow & [N/x]M \\
(\eta) \ \ \ \ \ \ \ \ \ \ \ [x:A]M(x) & \rightarrow  &M\ \  \ \ \ \ \ \ \ \  \ \ \  (x \notin FV(M)) \\
(\pi_1) \ \ \ \ [\langle r, l=a \colon A \rangle]  & \rightarrow  & r \\
(\pi_2) \ \ \ \ \langle r, l=a \colon A \rangle .l  & \rightarrow  & a \\
(\pi_2') \ \ \ \langle r, l=a \colon A \rangle.l'  & \rightarrow  & r.l' \ \ \  \ \ \ \ \ \ \ \ \ \ (l \neq l')
%%%%%%% May 10th remove %%%%%%%
%(\eta_{RCD}) \ \ \ \ & \langle [r], l=r.l \colon A \rangle \rightarrow_{ETARCD} \ r &
\end{eqnarray*}
We write $\rightarrow^+$ and $\rightarrow^*$ for the corresponding transitive closure and reflexive and transitive closure, respectively.

A term of the form on the left of an arrow is called a \emph{redex}.  For example, a $\pi_2$-redex is a term of the form $\langle r, l=a \colon A \rangle .l$.
\end{defn}

\selfc{
\begin{definition}[Compatible Closure] \label{comp-clos}
Let $R$ be a relation on terms.   The \emph{compatible closure} of $R$, notation $\triangleright_R$, is the least relation satisfying the following rules:
$$
\frac{M\ R\ N}{M\ \triangleright_R\ N} (R-Inc) \ \ \
\frac{A_1\ \triangleright_R\ B_1}{(x:A_1)A_2 \ \triangleright_R\ (x:B_1)A_2} (\Pi-L) \ \ \
\frac{A_2\ \triangleright_R\ B_2}{(x:A_1)A_2 \ \triangleright_R\ (x:A_1)B_2} (\Pi-R)
$$
$$
\frac{M\ \triangleright_R\ N}{El(M)\ \triangleright_R\ El(N)} (El) \ \ \
\frac{R_1\ \triangleright_R\ R_2}{\langle R_1, l \colon A\rangle\ \triangleright_R\ \langle R_2, l \colon A \rangle} (RCD_{fst}) \ \ \
\frac{A_1\ \triangleright_R\ B_1}{\langle R, l \colon A_1\rangle\ \triangleright_R\ \langle R, l \colon B_1 \rangle} (RCD_{snd}) \ \ \
$$
$$
\frac{A_1\ \triangleright_R\ B_1}{[x:A_1]M_0 \ \triangleright_R\ [x:B_1]M_0} (\lambda-L) \ \ \
\frac{M\ \triangleright_R\ N}{[x:A]M\ \triangleright_R\ [x:A]N} (\xi) \ \ \
\frac{A_1\ \triangleright_R\ B_1}{\langle r, l=a \colon A_1\rangle\ \triangleright_R\ \langle r, l=a \colon B_1\rangle} (rcd)
$$
$$
\frac{r_1\ \triangleright_R\ r_2}{\langle r_1, l=a \colon A\rangle\ \triangleright_R\ \langle r_2, l=a \colon A\rangle} (rcd_{fst}) \ \ \
\frac{a_1\ \triangleright_R\ a_2}{\langle r, l=a_1 \colon A\rangle\ \triangleright_R\ \langle r, l=a_2 \colon A\rangle} (rcd_{snd})
%\frac{M\ \triangleright_R\ N}{\langle M, l = a \colon  A \rangle\ \triangleright_R\ \langle N, l = a \colon A \rangle} (\xi_{rcd})
$$
$$
\frac{M\ \triangleright_R\ N}{M(P)\ \triangleright_R\ N(P)} (App-L) \ \ \
\frac{M\ \triangleright_R\ N}{P(M)\ \triangleright_R\ P(N)} (App-R) \ \ \
\frac{M\ \triangleright_R\ N}{M.l\ \triangleright_R\ N.l} (Sel) \ \ \
\frac{M\ \triangleright_R\ N}{[M]\ \triangleright_R\ [N]} (Res) \ \ \
$$
\\
\end{definition}

%\begin{definition}[Strongly Normalizable]
%\ \\
%A term in the system IDRT is strongly normalizable iff every reduction sequence starting from this term is finite. \\
%(\emph{two ways of definitions: either direct or inductively}). \\
%Strong normalisation for types, written $SN(A)$, is the least predicate closed under the following rule of inference:
%$$
%SN-_{I} \ \ \frac{for\ all\ B.(A \rightarrow^* B) \Rightarrow SN(B)}{SN(A)}
%$$
%and similarly for kinds. \\
%\end{definition}
%\noindent
%Also we give the definition of untyped reduction with regard to dependent records. \\

\begin{definition}[Untyped Reduction $\rightarrow^*$] \label{un-red}
%\ \\
We introduce the one-step reduction rules over untyped terms in IDRT:
the untyped reduction (or just reduction) $\rightarrow$ is the compatible closure of all the following rules in Figure \ref{Untyped-reduction}; we write $\rightarrow^+$ for the transitive closure of reduction and $\rightarrow^*$ for the reflexive, transitive closure of reduction. We also write $\rightarrow_{R}$ for the compatible closure of the last three rules that operate only on the record terms, and $\rightarrow_{R}^+$ and $\rightarrow_{R}^*$ accordingly. Similarly $\rightarrow_{\beta R}$ is the compatible closure of $\rightarrow_\beta$ and $\rightarrow_R$, and $\rightarrow_{\beta R}^+$ and $\rightarrow_{\beta R}^*$ accordingly. \\
\begin{figure}[here]
\framebox[5.8in][c]{
\begin{minipage}{\linewidth}
\begin{eqnarray*}
(\beta) \ \ \ \ & ([x:A]M)N \rightarrow_\beta \ [N/x]M &\\
(\eta) \ \ \ \ & [x:A]M(x) \rightarrow_\eta \ M & (x \notin FV(M)) \\
(restriction) \ \ \ \ & [\langle r, l=a \colon A \rangle] \rightarrow_{RESTR} \ r &\\
(field-selection) \ \ \ \ & \langle r, l=a \colon A \rangle .l \rightarrow_{FLDSEL} \ a &\\
(diff.\ field-sele.) \ \ \ \ & \langle r, l=a \colon A \rangle.l' \rightarrow_{FLDSL'} \ r.l' & (l \neq l') \\
%%%%%%% May 10th remove %%%%%%%
%(\eta_{RCD}) \ \ \ \ & \langle [r], l=r.l \colon A \rangle \rightarrow_{ETARCD} \ r & \\
\end{eqnarray*}
\end{minipage}
} \caption{One-step Reduction of Untyped Terms in IDRT} \label{Untyped-reduction}
\end{figure}
(\textbf{Redex}) Let $R$ be a relation, a term $M$ is an $R$-redex if there is some $N$ such that $M\ R\ N$. A term $M$ in IDRT is a redex if $M$ is a $\rightarrow$-redex. \\
\end{definition}

}%selfc

\begin{defn}[Weak-Head Normal Forms and Normal Forms] \label{WHN-N}\
\begin{itemize}
  \item A term $M$ is \emph{weak-head normal} if
  \begin{itemize}
    \item $M\equiv x$ is a variable;
    \item $M\equiv [x:K]k$;
    \item $M\equiv f(a)$, where $f$ is weak-head normal and not an abstraction;
    \item $M\equiv\record{}$;
    \item $M\equiv \record{r,\ l=a\colon A}$; or
    \item $M\equiv [r]$ or $M\equiv r.l$, where $r$ is weak-head normal and not a pair-record.
  \end{itemize}

  \item A term $M$ is \emph{normal} if
  \begin{itemize}
    \item $M\equiv x$ is a variable;
    \item $M\equiv [x:K]k$, which is not an $\eta$-redex, and $K$ and $k$ are normal;
    \item $M\equiv f(a)$ and $f$ and $a$ are normal and $f$ not an abstraction;
    \item $M\equiv \record{}$;
    \item $M\equiv \record{r,\ l=a\colon A}$ and $r$, $a$ and $A$ are normal.
    \item $M\equiv [r]$ or $M\equiv r.l$, where $r$ is normal and not a pair-record.
  \end{itemize}
\end{itemize}
The notions of weak-head normal forms and normal forms are lifted to record types, kinds and contexts in the usual way.
\end{defn}

This case was in our original proof of a DRT system with the WER rules, it was of interest because the weakly extensional rules are $\eta$-like rules that cause problems, such as strong normalization fails for untyped raw terms. For reason of discussion we keep this case still here.
The following lemmas show that the notion of computation captured in TOS is adequate \wrt\ the untyped reduction and the associated notions of normal forms.

\begin{lemma}[Adequacy of TOS \wrt\ Untyped Reduction] \label{AUR}\
\begin{itemize}
  \item If $\Gamma \models A \rightarrow C$ then there exists $B$ such that $A \rightarrow_{\beta R}^* B \rightarrow_\eta^* C$.
  \item If $\Gamma \models M \rightarrow N \rightarrow P \colon A$, then there exists $N'$ such that $M \rightarrow_{\beta R}^* N \rightarrow_{\beta R}^* N' \rightarrow_\eta^* P$.
\end{itemize}
\end{lemma}
\textbf{Proof.} By induction on derivations.

\begin{lemma}[Adequacy of TOS \wrt\ Normal Forms and WHNFs] \label{ANF}\
\begin{itemize}
  \item If $\models \Gamma \rightarrow \Delta$, then $\Delta$ is normal.
  \item If $\Gamma \models A \rightarrow B$, then $B$ is normal.
  \item If $\Gamma \models M \rightarrow N \rightarrow P \colon A$, then $N$ is weak-head normal and $P$ and $A$ are normal.
\end{itemize}
\end{lemma}
\textbf{Proof.} By induction on derivations.

\selfc{%moved as an earlier lemma (the last clause of the basic structural lemmas)
\begin{lemma}[Determinacy (Unique Normal Form)] \label{UNF}
%(Uniqueness of normal form) \\
\ \\
(1) If $\models \Gamma \rightarrow \Delta$, $\models \Gamma \rightarrow \Phi$, then $\Delta \equiv \Phi$; \\
(2) If $\Gamma \models A \rightarrow B$, $\Gamma \models A \rightarrow C$, then $B \equiv C$; \\
(3) If $\Gamma \models M \rightarrow N \rightarrow P \colon B$, $\Gamma \models M \rightarrow Q \rightarrow R \colon C$, then $N \equiv Q$, $P \equiv R$, $B \equiv C$.
\end{lemma}
%\begin{proof}
\textbf{Proof.} By simultaneous induction on derivations.
%\end{proof}
}%selfc

\subsubsection{Soundness and Completeness}
\label{subsec:comp-sound}

The TOS we have studied is sound and complete \wrt\ the type system IDRT of dependent record types.  In the informal introduction to TOS in Section~\ref{sec:introTOS}, we have over-simplified the situation.  In fact, what we shall do is to show that completeness holds for a simpler system $IDRT^-$ (with judgements of the form $\G\ts^- J$), which is obtained from IDRT by removing the seven substitution rules in Appendix~\ref{app:LF-rules}.  Therefore, the soundness and completeness may be pictured as follows:

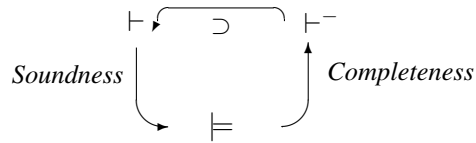
\begin{picture}(100, 65)(-115, 0)
\put(50, 50){\makebox(0,0){$\vdash$}}
\put(120, 50){\makebox(0,0){$\vdash^-$}}
\put(82, 10){\makebox(0,0){$\models$}}
\put(60, 40){\oval(20, 60)[bl]}
\put(105, 40){\oval(20, 60)[br]}
\put(82, 50){\oval(48, 10)[t]}
\put(24, 30){\makebox(0,0){\emph{\begin{small}Soundness\end{small}}}}
\put(150, 30){\makebox(0,0){\emph{\begin{small}Completeness\end{small}}}}
\put(82, 48){\makebox(0,0){\emph{\begin{small}$\supset$\end{small}}}}
\put(60, 10){\vector(1, 0){2}}
\put(115, 40){\vector(0, 1){2}}
\put(58, 50){\vector(-1, -2){2}}
\end{picture}

\begin{theorem}[Completeness of TOS \wrt\ $IDRT^-$] \label{compl}\
\begin{itemize}
  \item If $\Gamma \models ok$ then $\vdash^- \Gamma \ valid$.
  \item If $\Gamma \models A \rightarrow B$ then $\Gamma \vdash^- A \ kind$ and $\Gamma \vdash^- A=B$.
  \item If $\Gamma \models M \rightarrow N \rightarrow P \colon A$ then $\Gamma \vdash^- M \colon A$, $\Gamma \vdash^- M=N \colon A$, $\Gamma \vdash^- M=P \colon A$ and $\Gamma \vdash^- A=A$.
\end{itemize}
\end{theorem}
\textbf{Proof.} By simultaneous induction on derivations and examining each case of the $TOS$ inference rules.

\begin{cor}[Completeness of TOS \wrt\ IDRT] \label{completeness}\
\begin{itemize}
  \item If $\Gamma \models ok$ then $\vdash \Gamma \ valid$.
  \item If $\Gamma \models A \rightarrow B$ then $\Gamma \vdash A \ kind$ and $\Gamma \vdash A=B$.
  \item If $\Gamma \models M \rightarrow N \rightarrow P \colon A$ then $\Gamma \vdash M \colon A$, $\Gamma \vdash M=N \colon A$, $\Gamma \vdash M=P \colon A$ and $\Gamma \vdash A=A$.
\end{itemize}
\end{cor}
%\begin{proof}
\textbf{Proof.} By Theorem~\ref{compl} and the inclusion of $IDRT^-$ in IDRT.

\ \\

The Soundness Theorem is harder to prove.  We have to consider all the inference rules of IDRT including the structural rules.   In the following, we only consider some selected cases.  The detailed proof can be found in \cite{Fen10}.

\begin{theorem}[Soundness of TOS \wrt\ IDRT] \label{Soundness-TOS-LF}\
\begin{itemize}
  \item If $\Gamma \vdash ok$, then there exists $\Delta$ such that $\models \Gamma \rightarrow \Delta$.
  \item If $\Gamma \vdash A\ kind$, then there exists $B$ such that $\Gamma \models A \rightarrow B$.
  \item If $\Gamma \vdash A=B$ then there exists $C$ such that $\Gamma \models A \rightarrow C$ and $\Gamma \models B \rightarrow C$.
  \item If $\Gamma \vdash M \colon A$ then there exist $P$, $B$ such that $\Gamma \models A \rightarrow B$ and $\Gamma \models M \rightarrow_n P \colon B$.
  \item If $\Gamma \vdash M=N \colon A$, then there exist $P$, $B$ such that $\Gamma \models A \rightarrow B$, and $\Gamma \models M \rightarrow_n P \colon B$, $\Gamma \models N \rightarrow_n P \colon B$.
\end{itemize}
\end{theorem}
\textbf{Proof.} By induction on derivations. For the cases of LF-rules, see \cite{healf:TLCA99YY}.  We consider the following two cases about record types.
\begin{itemize}
  \item The second introduction rule in Figure~\ref{DRT-rules}:
  \[ \frac{\Gamma \vdash \langle R, \ l\colon A \rangle \colon RType \: \: \: \Gamma \vdash r \colon R \: \: \: \Gamma \vdash a \colon A(r)}{\Gamma \vdash \langle r, \ l=a \colon A \rangle \colon \langle R, \ l \colon A \rangle}
  \]
  By induction hypothesis, the following hold:
  \begin{enumerate}
    \item \label{(*)}
      $\Gamma \models \langle R, l \colon A \rangle \rightarrow_n \langle P, l \colon B \rangle \colon RType$ for some $P$ and $B$,
    \item $\Gamma \models r \rightarrow_n p \colon P'$ and $\Gamma \models R \rightarrow_n P' \colon RType[L]$ for some $p$, $P'$ and $L$, and
    \item \label{(**)}
      $\Gamma \models a \rightarrow_n b \colon C$ and $\Gamma \models A(r) \rightarrow C$  for some $b$ and $C$.
  \end{enumerate}
  By Lemma~\ref{Stru-prop}(\ref{UNF}) (Determinacy) and inversion of the rule $(RCDT)$ in Figure~\ref{TOS-DRT-rules}, $P \equiv P'$.  Therefore, by rule $(RCD)$ in Figure~\ref{TOS-DRT-rules}, $\Gamma \models \langle r, l=a \colon A \rangle \rightarrow_n \langle p, l=b \colon B \rangle \colon \langle P, l \colon B \rangle$.

  \item The third elimination rule in Figure~\ref{DRT-rules}:
  \[ \frac{\Gamma \vdash r \colon \langle R, \ l \colon A \rangle \: \: \: \Gamma \vdash [r].l' \colon B \: \: \: l \neq l'}{\Gamma \vdash r.l' \colon B}
  \]
  By induction hypothesis, the following hold:
  \begin{enumerate}
    \item $\Gamma \models r \rightarrow_n s \colon \langle P, l \colon B \rangle$ and $\Gamma \models \langle R, l \colon A \rangle \rightarrow \langle P, l \colon B \rangle$ for some $s$, $P$ and $B$, and
    \item $\Gamma \models [r].l' \rightarrow_n c \colon C$ and $\Gamma \models B \rightarrow C$ for some $c$ and $C$.
  \end{enumerate}
  Since $l \neq l'$, by $(FLDSL')$ in Figure~\ref{TOS-DRT-rules}, we have $\Gamma \models r.l' \rightarrow_n c \colon C$.
\end{itemize}

\selfc{
We shall examine each case of $LF$ and $DRT$ inference rules
%in Figure \ref{TOS-LF-rules} and \ref{TOS-DRT-rules}
and consider all the structural rules (selected proof segment): \\
-- Kinding rules. We show for $RType$: By induction hypothesis $\Gamma \models ok$, using $RTYPE_{TOS}$ we have $\Gamma \models RType \rightarrow RType$. \\
-- Formation rules for $RType[\_]$. We show for the second one: By induction hypothesis there exist $P, B'$ such that $\Gamma \models RType[L] \rightarrow B'$ (so by inversion $B' \equiv RType[L]$), and $\Gamma \models R \rightarrow_n P \colon RType[L]$. Also by induction hypothesis there exist $B$ such that $\Gamma \models A \rightarrow B \colon (P)Type$. Together with the condition $l \notin L$, using $RCDT_{TOS}$ we have $\Gamma \models \langle R, l \colon A \rangle \rightarrow_n \langle P, l \colon B \rangle \colon RType[L \cup \{l\}]$. \\
-- Introduction rules. We show for the second one: By induction hypothesis there exist $P, B$ such that $\Gamma \models \langle R, l \colon A \rangle \rightarrow_n \langle P, l \colon B \rangle \colon RType$ $^{(*)}$; also that there exist $s, P'$ such that $\Gamma \models r \rightarrow_n s \colon P'$ and $\Gamma \models R \rightarrow_n P' \colon RType[L]$ for some $L$. By inversion of $RCDT_{TOS}$ and by Determinacy we know $P \equiv P'$. Also by induction hypothesis there exist $b, B'$ such that $\Gamma \models a \rightarrow_n b \colon B'$ and $\Gamma \models A(r) \rightarrow B'$ $^{(**)}$. Here, consider two subcases when we construct the new normal form for the small record $\langle r, l=a \colon A \rangle$: \\
\indent
First subcase, $\langle s, l=b \colon B \rangle$ is not an $\eta$-redex. By inversion on $^{(*)}$, we have $\Gamma \models A \rightarrow_n B \colon (P)Type$ and $\Gamma \models r \rightarrow_n s \colon P$. Use the $TOS-LF$ rule for application, we have $\Gamma \models A(r) \rightarrow_n B(s) \colon Type$. Use the $TOS-LF$ rule for $El$, we have $\Gamma \models El(A(r)) \rightarrow El(B(s))$ $^{(***)}$. Compare $^{(**)}$ and $^{(***)}$, by Determinacy we have $El(B(s)) \equiv El(B')$. By inversion on $LF$'s $El$ rule we have $\Gamma \vdash B(s) = B' \colon Type$. Apply induction hypothesis again and by the fact that $B'$ is normal (Adequacy for normal forms, Lemma \ref{ANF}), we have $\Gamma \models B(s) \rightarrow_n B' \colon Type$. Finally, with all these consitions got, we apply $RCD_{TOS}$ rule and get $\Gamma \models \langle r, l=a \colon A \rangle \rightarrow_n \langle s, l=b \colon B \rangle \colon \langle P, l \colon B \rangle$. \\
\indent
Second subcase, $\langle s, l=b \colon B \rangle$ is an $\eta$-redex. This means there exists $r'$ such that $[r'] \equiv s$ and $r'.l \equiv b$. Similarly to the above subcase, by inversion on $BASE_{RESTR}$ and $BASE_{FLDSEL}$ rules of $TOS-DRT$, we have $\Gamma \models r' \rightarrow_n p \colon \langle P, l \colon B \rangle$ for some $p$. \\
%WRONG==By Determinacy we could have $B \equiv B'$(?). Thus using $RCD_{TOS}$ we have $\Gamma \models \langle r, l = a \colon A \rangle \rightarrow_n \langle s, l = b \colon B \rangle \colon \langle P, l \colon B \rangle$. \\
%\\
-- Elimination rules. We show for the third one (the first two could be easily deduced from inverting the $BASE_{TOS}$ rules): By induction hypothesis there exist $p, b, P, B$ such that $\Gamma \models r \rightarrow_n \langle p, l = b \colon B \rangle \colon \langle P, l \colon B \rangle$ and $\Gamma \models \langle R, l \colon A \rangle \rightarrow \langle P, l \colon B \rangle$; also that there exist $c$ such that $\Gamma \models [r].l' \rightarrow_n c \colon C$ and $\Gamma \models B \rightarrow C$. There are two subcases here, apply an inversion either on the rule $FLDSEL_{TOS}$ or on the rule $BASE_{TOS}$, we will have $\Gamma \models B(P) \rightarrow C \colon Type$. Together with the condition $l \neq l'$, using $FLDSL'_{TOS}$ we have $\Gamma \models r.l' \rightarrow_n c \colon C$. \\
%\\
-- Computation rules. We show for the third one: From the induction hypothesis and by inversion on the rule $RESTR_{TOS}$ and using the rule $FLDSL'_{TOS}$ it will be easily derived that $\Gamma \models r.l' \rightarrow_n s \colon C$ and $\Gamma \models [r].l' \rightarrow_n s \colon C$ and $\Gamma \models B \rightarrow C$ for some $s, C$. %\\
-- Congruence rules. We show for the one for $RType[L \cup \{l\}]$: Using induction hypothesis and the rule $RCDT_{TOS}$, also by Determinacy, we will derive that $\Gamma \models \langle R, l \colon A \rangle \rightarrow \langle P, l \colon B \rangle \colon RType[L \cup \{l\}]$, $\Gamma \models \langle R', l \colon A' \rangle \rightarrow \langle P, l \colon B \rangle \colon RType[L \cup \{l\}]$ for some $P, B$ and $\Gamma \models RType[L \cup \{l\}] \rightarrow RType[L \cup \{l\}]$.
\qed \\
%\end{proof}

}%selfc

\subsection{Strong Normalisation in TOS}
\label{sec:PSR-SN}

The strong normalisation property of the TOS says that, if a term $M$ is well-typed in the TOS (\ie, $\G\models M\colon A$), then $M$ is strongly normalisable.  This result, together with the soundness theorem, will then be enough to show that the original type theory has the property of strong normalisation.

The strong normalisation property of the TOS by introducing a notion of \emph{parallel reduction} and showing that it has the so-called \emph{Parallel Subject Reduction} property.

\begin{definition}[Parallel Reduction] \label{Pal-Red}
Parallel reduction $\Rightarrow$ is defined as the least relation closed under the rules in Figure~\ref{Parallel-reduction}, and is extended in an obvious way to kinds and contexts.

\begin{figure}[here]
\framebox[5.8in][c]{
\begin{minipage}{\linewidth}
%$$for\ LF:$$
\ \\
$$
VAR\ \frac{}{x \Rightarrow x} \ \ \
\lambda\ \frac{A \Rightarrow A' \ \ M \Rightarrow M'}{[x:A]M \Rightarrow [x:A']M'} \ \ \
APP\ \frac{M \Rightarrow M' \ \ N \Rightarrow N'}{M(N) \Rightarrow M'(N')}
$$
$$
\beta\ \frac{M \Rightarrow M' \ \ N \Rightarrow N'}{([x:A]M)(N) \Rightarrow [N'/x]M'} \ \ \
\eta\ \frac{M_0 \Rightarrow M'(x) \ \ x \notin FV(M')}{[x:A_1]M_0 \Rightarrow M'}
$$
%$$for\ DRT:$$
$$
RCDT_{EMP}\ \frac{}{\langle \rangle \Rightarrow \langle \rangle} \ \ \ \ \
RCDT\ \frac{R \Rightarrow R' \ \ A \Rightarrow A'}{\langle R, l \colon A \rangle \Rightarrow \langle R', l \colon A' \rangle}
$$
$$
RCD_{EMP}\ \frac{}{\langle \rangle \Rightarrow \langle \rangle} \ \ \ \
RCD\ \frac{r \Rightarrow r' \ \ a \Rightarrow a' \ \ A \Rightarrow A'}{\langle r, l = a \colon A \rangle \Rightarrow \langle r', l = a' \colon A' \rangle}
$$
$$
BASE_{RESTR}\ \frac{r \Rightarrow r'}{[r] \Rightarrow [r']}\ \ \ \
BASE_{FLDSEL}\ \frac{r \Rightarrow r'}{r.l \Rightarrow r'.l}
$$
$$
RESTR\ \frac{r \Rightarrow r'}{[\langle r, l=a \colon A \rangle] \Rightarrow r'} \ \ \ \
FLDSEL\ \frac{a \Rightarrow a'}{\langle r, l=a \colon A \rangle.l \Rightarrow a'}
$$
$$
FLDSL'\ \frac{r \Rightarrow r' \ \ l \neq l'}{\langle r, l=a \colon A \rangle.l' \Rightarrow r'.l' } \ \ \ \
%\eta RCD\ \frac{r \Rightarrow r'}{\langle [r], l=r.l \colon A \rangle \Rightarrow r'}
$$
\\
\end{minipage}
}%\framebox
\caption{Parallel Reduction for IDRT} \label{Parallel-reduction}
\end{figure}
%contexts shouldn't be included for reduction ? -- should, because in the lemma of para. sub. red. there are Gamma => Gamma' ! -- Sept 17
%(how to extend with the $DRT$ terms ?)--did on Sept 15 \\
\end{definition}

\remark\ Parallel reduction has some simple properties. First, $M\To M$ for all M. Furthermore, if $M\to N$ then $M\To N$, and if $M\To N$ then $M\to^\ast N$. Finally, if $M\To M'$ and $N\To N'$ then $[N/x]M \To [N'/x]M'$.

\selfc{
\begin{lemma}[Parallel Reduction $\Rightarrow$ $\subset$ Untyped Reduction $\rightarrow^*$] \label{rel-reductions}
%\ \\
If $M \Rightarrow N$ then $M \rightarrow^* N$. %\\
\end{lemma}
\textbf{Proof.} By structural induction on the terms. From Definition \ref{Parallel-reduction} we know that no $\rightarrow^*$-redex was created by parallel reduction $\Rightarrow$. \footnote{As consequence of Lemma \ref{rel-reductions}, we have $\rightarrow \ \ \subset \ \ \Rightarrow \ \ \subset \ \ \rightarrow^* \ \ \subset \ \ \Rightarrow^*$, Church-Rosser holds for both $\rightarrow^*$ and $\Rightarrow^*$. } \qed \\
%\noindent
}%selfc

\begin{lemma}[Parallel Subject Reduction] \label{PSR}\
\begin{enumerate}
  \item If $\models \Gamma \rightarrow \Delta$ and $\Gamma \Rightarrow \Gamma'$, then $\models \Gamma' \rightarrow \Delta$.
  \item If $\Gamma \models A \rightarrow B$, $\Gamma \Rightarrow \Gamma'$ and $A \Rightarrow A'$, then $\Gamma' \models A' \rightarrow B$.
  \item If $\Gamma \models M \rightarrow N \rightarrow P \colon A$, $\Gamma \Rightarrow \Gamma'$ and $M \Rightarrow M'$, then there exist $N'$ and $N''$ such that $N \Rightarrow N'$, $\Gamma' \models M' \rightarrow N'' \rightarrow P \colon A$ and $\Gamma' \models N' \rightarrow N'' \rightarrow P \colon A$.
\end{enumerate}
\end{lemma}
\textbf{Proof.} By simultaneous induction on derivations. The detailed proof can be found in \cite{Fen10}.

\begin{lemma}[Subject Reduction] \label{SR}
If $\Gamma \models M \rightarrow N \rightarrow P \colon A$, $\Gamma \Rightarrow \Gamma'$ and $M \Rightarrow M'$, then there exists $N'$ such that $\Gamma' \models M' \rightarrow N' \rightarrow P \colon A$ and $N \rightarrow^* N'$.
\end{lemma}
\textbf{Proof.} By simultaneous induction on derivations, using Lemma \ref{PSR}.
\ \\

The proof of strong normalisation of the TOS (Theorem~\ref{SN-TOS}) uses the following lemma.

\begin{lemma}\label{NOT-ABS-PAIR}\
\begin{itemize}
  \item If $M$ is weak-head normal and not an abstraction and $M \rightarrow^* N$, then $N$ is weak-head normal and not an abstraction.
  \item If $M$ is weak-head normal and not a pair record and $M \rightarrow^* N$, then $N$ is weak-head normal and not a pair record.
\end{itemize}
\end{lemma}
\textbf{Proof.} By induction on length of reduction for untyped terms.

\begin{theorem}[Strong Normalisation of TOS] \label{SN-TOS}\
\begin{enumerate}
  \item If $\Gamma \models A \rightarrow B$ then $A$ is strongly normalisable.
  \item If $\Gamma \models M \rightarrow N \rightarrow P \colon A$ then $M$ is strongly normalisable.
\end{enumerate}
\end{theorem}
\textbf{Proof.} By simultaneous induction on derivations.   The full proof can be found in \cite{Fen10}.  We shall consider one of the most difficult cases -- when the last rule used is $(BASE_{RESTR})$ in Figure~\ref{TOS-DRT-rules}:
  $$
    \frac{\Gamma \models r \rightarrow q \rightarrow p \colon \langle P,\ l \colon B \rangle \: \: \: \: p,\ q\ not\ pair\texttt{-}records}{\Gamma \models [r] \rightarrow [q] \rightarrow [p] \colon P}
  $$
By \IH, $r$ is \SN.  It suffices for us to show that $[r]$ is \SN\ if $\Gamma\models r \rightarrow_w q \colon \langle P, l \colon B \rangle$ such that $q$ is not a pair-record.  We do this by induction on the maximal length of reductions starting from $r$.

Assume that $r\to r_1$.  \selfc{As $q$ is not a pair-record and $r_1$ is arbitrary, it is enough to show that $[r_1]$ is \SN. Since $r\to r_1$,}  We then have $r\To r_1$, which implies by Lemma~\ref{PSR} (Parallel Subject Reduction) that there exist $r'$ and $r''$ such that
\[
\Gamma \models r_1 \rightarrow_w r'' \colon \langle P, l \colon B\rangle, \ \ \
\Gamma \models r' \rightarrow_w r'' \colon \langle P, l \colon B\rangle, \ \ \ \textrm{and}\ \ \
q \Rightarrow r'.
\]
We therefore have that $q$ is weak-head normal (by Lemma~\ref{ANF}) and that $q \rightarrow^* r'$ (see the remark above).  From these and Lemma~\ref{NOT-ABS-PAIR}, we have
\begin{itemize}
  \item[(*)] $r'$ is weak-head normal and not a pair-record.
\end{itemize}
Furthermore, by Lemma~\ref{AUR}, we have
\begin{itemize}
  \item[(**)] $r' \rightarrow^* r''$.
\end{itemize}
From $(*)$ and $(**)$, $r''$ is not a pair-record by Lemma~\ref{NOT-ABS-PAIR}.  Therefore, since $\Gamma \models r_1 \rightarrow_w r'' \colon \langle P, l \colon B\rangle$, we conclude by \IH\ that $[r_1]$ is \SN.

\selfc{
\par{And eventually, we arrive at the following main Corollary, which is strong normalisation of our original system of dependent record types IDRT: }

\begin{corollary}[Strong Normalisation of IDRT] \label{SN-LFDRT}
\ \\
%Presented as Proposition \ref{SN-LF}.-- Should be rephrased for contexts, kinds and terms.
(1) If $\Gamma\ valid$ then $\Gamma$ is strongly normalisable; \\
(2) If $\Gamma \vdash A \ kind$ then $A$ is strongly normalisable; \\
(3) If $\Gamma \vdash A = B$ then both $A$ and $B$ are strongly normalisable to some $C$; \\
(4) If $\Gamma \vdash M \colon A$, then $M$ is strongly normalisable to some $P$, $A$ is strongly normalisable to some $B$ such that $P \colon B$ in $\Gamma$; \\
(5) If $\Gamma \vdash M = N \colon A$, then both $M$ and $N$ are strongly normalisable to some $P$, $A$ is strongly normalisable to some $B$ such that $P \colon B$ in $\Gamma$.
\end{corollary}
%\begin{proof}
\textbf{Proof.} Derived from Theorem \ref{Soundness-TOS-LF}
%(Soundness of $TOS$ w.r.t. IDRT)
and Theorem \ref{SN-TOS}.  \\
%(Strong Normalisation of $TOS$)
%\end{proof}

\par{\noindent Other metatheory such as the Church-Rosser property, uniqueness of normal form, etc. could be derived in a similar way as strong normalisation and subject reduction are done. }

\begin{corollary}[Church-Rosser Property of IDRT] \label{CR-LFDRT}
\ \\
If $\Gamma \vdash M = N \colon A$, then $\exists P$ such that $M \rightarrow^* P$ and $N \rightarrow^* P$.
\end{corollary}
\textbf{Proof.} By Soundness (Theorem \ref{Soundness-TOS-LF}), there exist $P, B$ such that $\Gamma \models M \rightarrow_n P \colon B$ and $\Gamma \models N \rightarrow_n P \colon B$. By Adequacy (Lemma \ref{AUR}) we prove that $M \rightarrow^* P$ and $N \rightarrow^* P$. \\
}%selfc

\section{Meta-theoretic Properties of IDRT}
\label{sec:metaDRT}

From the properties of the TOS that have been proved in the last section, the meta-theoretic properties of IDRT, the type theory for dependent record types, can be proved.  Here, we give the theorems for Subject Reduction, Church-Rosser and Strong Normalisation.  (For further details and other properties, see \cite{Fen10}.)

\begin{thm}[Subject Reduction for IDRT] \label{SR-LFDRT}
If $\Gamma \vdash M \colon A$ and $M \rightarrow N$, then $\Gamma \vdash N \colon A$.
\end{thm}
\textbf{Proof.} First of all, we have $\Gamma \models M \colon A$ (by the Soundness Theorem \ref{Soundness-TOS-LF}) and $M \Rightarrow N$ (since $M \rightarrow N$).  Therefore, by Lemma~\ref{SR}, we have $\Gamma \models N \colon A$.  Now, by Completeness (Theorem~\ref{completeness}), $\Gamma \vdash N \colon A$.

\begin{thm}[Strong Normalisation for IDRT] \label{SN-LFDRT}\
\begin{enumerate}
  \item If $\Gamma\ valid$, then $\Gamma$ is strongly normalisable.
  \item If $\Gamma \vdash A \ kind$, then $A$ is strongly normalisable.
  \item If $\Gamma \vdash A = B$, then both $A$ and $B$ are strongly normalisable to some $C$.
  \item If $\Gamma \vdash M \colon A$, then $M$ and $A$ are strongly normalisable and $\G\ts P \colon B$, where $P$ and $B$ are the normal forms of $M$ and $A$, respectively.
  \item If $\Gamma \vdash M = N \colon A$, then both $M$ and $N$ are strongly normalisable to some $P$, $A$ is strongly normalisable to some $B$ such that $\G\ts P \colon B$.
\end{enumerate}
\end{thm}
\textbf{Proof.} By the Soundness Theorem~\ref{Soundness-TOS-LF}
%(Soundness of $TOS$ w.r.t. IDRT)
and Theorem \ref{SN-TOS}.
%(Strong Normalisation of $TOS$)

\begin{thm}[Church-Rosser for IDRT] \label{CR-LFDRT}
If $\Gamma \vdash M = N \colon A$, then $M \rightarrow^* P$ and $N \rightarrow^* P$ for some $P$.
\end{thm}
\textbf{Proof.} By Soundness (Theorem \ref{Soundness-TOS-LF}), there exist $P$ and $B$ such that $\Gamma \models M \rightarrow_n P \colon B$ and $\Gamma \models N \rightarrow_n P \colon B$.  Then, by Adequacy (Lemma \ref{AUR}), we have $M \rightarrow^* P$ and $N \rightarrow^* P$.

\section{Conclusions}
\label{sec:conclusion}

We have studied the meta-theory of a type theory with dependent record types, by studying its typed operational semantics.  As we have mentioned in Section~\ref{sec:LFwithDRT}, dependent record \emph{types} are rather different from dependent record \emph{kinds}, with the former having a much richer structure and being more difficult to study.  The meta-theory of dependent record kinds has been studied by Coquand \emph{et. al} \cite{ctp:semantic-records05}, where they have given a proof of termination of type-checking.  As far as we know, ours is the first attempt to study the meta-theory of dependent record types formulated in a logical framework \cite{Pollack:records02,luo:TYPES08,luo:MLPA09}.  In the light that the TOS-approach has been successfully applied to the meta-theoretic study of type theories with $\eta$-equality \cite{healf:TLCA99YY} and with inductive types \cite{healf:thesis}, the current work can be used to justify the incorporation of dependent record types in a full-scale type theory as implemented in the proof assistants such as Agda and Coq.

The dependent record types studied in this paper are \emph{intensional} in the sense that we do not have the following extensional equality rules \cite{bet-tar:subtyping98,luo:TYPES08}:
$$
\frac{\Gamma \vdash r\colon \langle \rangle}{\Gamma \vdash r = \langle \rangle \colon \langle \rangle}
\: \: \: \: \: \:
\frac{\begin{array}{c} \Gamma \vdash r \colon \langle R, \ l\colon A \rangle \: \: \: \Gamma \vdash r' \colon \langle R, \ l\colon A \rangle \\ \Gamma \vdash [r] = [r']\colon R \: \: \: \Gamma \vdash r.l = r'.l \colon A([r]) \end{array}}{\Gamma \vdash r = r' \colon \langle R, \ l\colon A \rangle}
$$
They basically say that two records are computationally equal if their components are.  For instance, from the second rule above, we would have $\record{r,\ l=r.l} = r$ for any $r$ of type $\record{R,\ l\colon A}$.  It is unclear whether the TOS-approach as adopted in this paper can be applied to such (weakly) extensional record types.  It would be obviously problematic if one considered the reduction relation for the records as follows:
\[ \record{r,\ l=r.l} \to r \]
for, together with the $\eta$-reduction for $\lam$-terms, the Church-Rosser property would fail to hold.  A natural question arises here: would it possible if one takes the TOS-approach by considering a reduction relation that treats $\eta$-long normal forms (\eg, by taking the above reduction in the other direction)?  This involves the development of the TOS-approach to incorporate $\eta$-long normal forms and research is needed to see whether it is possible.

\selfc{

\par{
%The weakly extensional equality rules are the $\eta$-rules for the dependent record types, which means our system of DRTs allow $\eta$-rules, and the Church-Rosser property, i.e., two distinct reductions starting from the same term commutes, could fail.
This is generally problematic with systems with $\eta$-equality, an example for why Church-Rosser fails could be as: In a system with both $\Pi$-type, $\Sigma$-type and unit type $\textbf{1}$, suppose we allow $\eta$-rules for $\Pi$, $\Sigma$ and for unit types, }

\begin{eqnarray*}
%(\eta_1) \ \ \ x \rightarrow * \colon \textbf{1} \\
(\eta_\Pi) \ \ \ & [x:A]f(x) \ \rightarrow \ f \colon B \ (if\ x\ \notin\ FV(f)) \\
(\eta_\Sigma) \ \ \ & \langle \pi_1(x), \pi_2(x) \rangle \ \rightarrow \ x \colon A \\
(\eta_\textbf{1}) \ \ \ & x \ \rightarrow \ * \colon \textbf{1}
\end{eqnarray*}
then, a term $[x:A]f(x)$ with $f \colon A \rightarrow \textbf{1}$ could have two reductions by $(\eta_\Pi)$ and by $(\eta_\textbf{1})$ that do not commute: $[x:A]f(x)\ \rightarrow_{\eta_\Pi} \ f$ and $[x:A]f(x)\ \rightarrow_{\eta_\textbf{1}} \ [x:A].*$. Note that $[x:A].*$ is already a canonical form.
\par{Similarly for a term $\langle \pi_1(x), \pi_2(x) \rangle \colon \textbf{1} \times A$, it reduces to both $x$ and to $\langle *, \pi_2(x) \rangle$} under $(\eta_\Sigma)$ and $(\eta_\textbf{1})$ respectively, which do not commute.

This situation happens because there are two types of $\eta$-reductions: one is the structural $\eta$s such as $(\eta_\Pi)$ or $(\eta_\Sigma)$, the other one is the non-structural or ``terminating'' $\eta$s such as $(\eta_\textbf{1})$ (they are called ``terminating'' because the calculations are towards a canonical form). The mixture of these two different types of $\eta$-rules has caused the failure of the Church-Rosser property. But if a system simply has only the structural $\eta$s, Church-Rosser still holds, for example,

\begin{eqnarray*}
\ [x:A]\langle \pi_1(x), \pi_2(x) \rangle & \rightarrow_{\eta_\Pi} \ & \langle \pi_1(x), \pi_2(x) \rangle \ \  \rightarrow_{\eta_\Sigma} \ \ x, \ and \\
\ [x:A]\langle \pi_1(x), \pi_2(x) \rangle & \rightarrow_{\eta_\Sigma} \ & [x:A]x \ \ \rightarrow_{\eta_\Pi} \ \ x \\
\end{eqnarray*}

\par{In our system, the weakly extensional equality rules are similar to the $\eta$-rules, and in particular, if one allows $\eta$-reduction for the DRTs in the definition of untyped reduction (Definition \ref{Untyped-reduction}), our proofs will not work, because the first weakly extensional equality rule is one in the category of non-structural $\eta$s, while the second still belongs to the category of structural $\eta$s. So if this system is extended with non-structural $\eta$s, i.e., the first WER, Church-Rosser fails. However, in our proof, the part related to the weakly extensionality rules in TOS and in other related definitions are supposed to be \emph{disconnected} with other part, which means, if we add a case in the proof for Strong Normalisation that deals with the weakly extensional equality rules given above, %in Figure \ref{DRT-rules}
it will only use rules provided by the TOS rule $ETA_{RCD}$ in Figure \ref{TOS-DRT-rules} and the $\eta$-parallel reduction definition in Figure \ref{Parallel-reduction}. }
\par{In another word, if one takes the weakly extensional equality rules out of the system, and remove also the $ETA_{RCD}$ rule in TOS and the other related definitions, such as the ETA for parallel reduction, the proof still go through. If one adds back these rules concerning structural $\eta$-calculation, as a corollary, Church-Rosser works for the typeable terms in TOS. This typeable semantic control ensures that $\eta$s do not cause problem, and it is one of the main features of applying the TOS approach. This approach could be useful for other systems without $\eta$-definitions or non-structural $\eta$-definitions for similar metatheoretical proof. }
% to apply the similar idea of $TOS$

%\newpage

%Comparison to other SN proofs, more reason of using TOS. \\

\par{In this paper we have introduced a new formulation of dependent record types as an extension of the logical framework LF, and we give a typed operational semantics for this formulation. We have proved some metatheoretical results on this typed semantics, and by its soundness theorem, we have derived the strong normalisation of the dependent record types. Other metatheoretical results such as the Church-Rosser property, subject reduction, etc., could be similarly derived. }

\par{Coquand \emph{et. al} have presented also a metatheoretical study in \cite{ctp:semantic-records05}, on a system of record kinds. They have given a proof of termination of type-checking for the record types; different with their method, we have shown that the typed semantical method to prove strong normalisation that Goguen suggested could also apply on our dependent record types. }

\par{As Goguen's work has established a full metatheory for the UTT and the LF, we here also similarly set up a full metatheory for the DRTs with a same way. Our work has led to an understanding on the following two points: First, the DRTs with intensional manifest fields developed in previous work \cite{luo:TYPES08,luo:MLPA09} own good metatheoretical properties, which would support themselves further as a modelling foundation to module mechanism; Second, this technique used here of the typed operational semantics could be applied to other type systems with structural $\eta$-rules. In conclusion, the methodology of applying TOS has set up a relation between well-formedness in the TOS and well-behaveness in original systems, which is ensured by the typing control of the TOS. }

\par{In the future, we have a few attempts concerning the approach discussed in this paper. Firstly, we would like to extend this technique to wider range of type systems, such as to the Logical Framework with inductive schemata and the dependent record types together with other types such as the unit types.
%but in the proofs of metatheory there, pure proof-theoretic proofs are hard to achieve. i.e., without any model interpretation. Thus, this work could be seen as a simple yet nontrivial application of Goguen's technique which succeeds in ensuring the good metatheory of dependent record types. }
Also, we are interested to find out if this technique could be applied
%canonical normalisation of formal calculi, that how $TOS$ succeeds to capture canonical normalisation reductions via the typed control, and thus provided the purely proof-theoretic approach; however,
to ill-behaving (e.g. non-SN) calculi, or, for not only proving strong normalisation, we would like to also prove weak-head normalisation in related type systems. }
% we would like to know whether or not $TOS$ could also work, and if not, we'd also like to know why $TOS$ fails to work for them. }

%\par{On the other hand, on the non-canonical normalisation of formal calculi, we are interested to find out if $TOS$ works also for the weak-head normalisation. These are the open problems that wait for our furture work. }

\paragraph{Comparason with other Systems of Dependent Records}
\par{Dependent record types have been previously studied in \cite{MacQ:module86, Harper-Lillibridge93, bet-tar:subtyping98, Pollack:records02, ctp:semantic-records05}, with applications to the study of module mechanisms for both programming and proof languages.
%aiming at providing dependently typed means for modular structures such as module systems in functional programming languages and in proof languages.
MacQueen has firstly suggested in \cite{MacQ:module86} a module system using dependent $\Sigma$-types as module signatures, which combines the first ideas coming from ML on data abstraction, and from Pebble on dependent Sigma types.
Later, Harper and Lillibridge have put more emphasis on the \emph{control of information flow} between program units, i.e. the idea of sharing, and have proposed in \cite{Harper-Lillibridge93} the weak sum type to implement the ML-style ``sharing by equation'', or called ``sharing by coherence conditions''; but in their formulation, the bounded quantification used is proven to be undecidable, the reason is by the FORGET-rule which loses type information. This has been proved similarly as in Pierce's proof for undecidability of $F_{\leq}$ \cite{Pie94}.  }

\par{Betarte and Tasistro have suggested in \cite{bet-tar:subtyping98} a different record system, i.e. of \emph{record kinds},
%using setoid to represent record kinds as an extension of Martin-L\"{o}f's Logical Framework,
with type inclusion to represent subtyping, this system is proved to be decidable (the proof is called unicity of right identity); however the record kinds are simpler than record types as a structure.
Pollack presented in \cite{Pollack:records02} a system with dependent record types which is similar to ours however allows repetition of field labels. }

%for SN: We'd like to know how TOS succeeds to capture canonical normalisation reductions via typed control ; We'd also like to know why TOS fails to work for ill-behaving (e.g. non-SN) calculi
%for WHN: We'd like to know whether TOS works also for WHN (non-canonical)

}%selfc

\vspace{0.6cm}

\noindent\textbf{Acknowledgement} The authors would like to thank Robin Adams for discussions on dependent record types and the first author thanks Cody Roux for discussions.

\bibliographystyle{alpha}
\bibliography{bib}

\selfc{

}%selfc

%\newpage
\appendix

\section{Inference Rules of LF}
\label{app:LF-rules}

The inference rules of the logical framework LF are given below.  (See Chapter 9 of \cite{luo:book94} for further details.)

\ \\
\noindent
\ \ \ \emph{Contexts and assumptions}
$$
% empty context I change it from <> to ()  --
\frac{}{ () \: \: valid}\: \: \: \frac{\Gamma \vdash K\:
\: kind\: \: \: x\notin FV(\Gamma )}{\Gamma ,x:K\: \: valid}\: \: \:
\frac{\Gamma ,x:K,\Gamma '\: \: valid}{\Gamma ,x:K,\Gamma '\vdash
x\colon K}$$
\emph{General equality rules}
$$
\frac{\Gamma \vdash K\: \: kind}{\Gamma \vdash K=K}\: \: \:
\frac{\Gamma \vdash K=K'}{\Gamma \vdash K'=K}\: \: \: \frac{\Gamma
\vdash K=K'\: \: \Gamma \vdash K'=K''}{\Gamma \vdash K=K''}$$
$$
\frac{\Gamma \vdash k\colon K}{\Gamma \vdash k=k\colon K}\: \: \:
\frac{\Gamma \vdash k=k'\colon K}{\Gamma \vdash k'=k\colon K}\: \:
\: \frac{\Gamma \vdash k=k'\colon K\: \: \Gamma \vdash k'=k''\colon
K}{\Gamma \vdash k=k''\colon K}$$
\emph{Equality typing rules}
$$
\frac{\Gamma \vdash k\colon K\: \: \Gamma \vdash K=K'}{\Gamma \vdash
k\colon K'}\: \: \: \frac{\Gamma \vdash k=k'\colon K\: \: \Gamma
\vdash K=K'}{\Gamma \vdash k=k'\colon K'}$$
\emph{Substitution rules}
$$
\frac{\Gamma ,x:K,\Gamma '\: \: valid\: \: \Gamma \vdash k\colon
K}{\Gamma ,[k/x]\Gamma '\: \: valid}$$
$$
\frac{\Gamma ,x:K,\Gamma '\vdash K'\: \: kind\: \: \Gamma \vdash
k\colon K}{\Gamma ,[k/x]\Gamma '\vdash [k/x]K'\: \: kind}\: \: \:
\frac{\Gamma ,x:K,\Gamma '\vdash K'\: \: kind\: \: \Gamma \vdash
k=k'\colon K}{\Gamma ,[k/x]\Gamma '\vdash [k/x]K'=[k'/x]K'}$$
$$
\frac{\Gamma ,x:K,\Gamma '\vdash k'\colon K'\: \: \Gamma \vdash
k\colon K}{\Gamma ,[k/x]\Gamma '\vdash [k/x]k'\colon [k/x]K'}\: \:
\: \frac{\Gamma ,x:K,\Gamma '\vdash k'\colon K'\: \: \Gamma \vdash
k_{1}=k_{2}\colon K}{\Gamma ,[k_{1}/x]\Gamma '\vdash
[k_{1}/x]k'=[k_{2}/x]k'\colon [k_{1}/x]K'}$$
$$
\frac{\Gamma ,x:K,\Gamma '\vdash K'=K''\: \: \Gamma \vdash k\colon
K}{\Gamma ,[k/x]\Gamma '\vdash [k/x]K'=[k/x]K''}\: \: \:
\frac{\Gamma ,x:K,\Gamma '\vdash k'=k''\colon K'\: \: \Gamma \vdash
k\colon K}{\Gamma ,[k/x]\Gamma '\vdash [k/x]k'=[k/x]k''\colon
[k/x]K'}$$
\emph{The kind Type}
$$
\frac{\Gamma \: \: valid}{\Gamma \vdash Type\: \: kind}\: \: \:
\frac{\Gamma \vdash A\colon Type}{\Gamma \vdash El(A)\: \: kind}\:
\: \: \frac{\Gamma \vdash A=B\colon Type}{\Gamma \vdash
El(A)=El(B)}$$
\emph{Dependent product kinds}
$$
\frac{\Gamma \vdash K\: \: kind\: \: \Gamma ,x:K\vdash K'\: \:
kind}{\Gamma \vdash (x:K)K'\: \: kind}\: \: \: \frac{\Gamma \vdash
K_{1}=K_{2}\: \: \Gamma ,x:K_{1}\vdash K_{1}'=K_{2}'}{\Gamma \vdash
(x:K_{1})K_{1}'=(x:K_{2})K_{2}'}$$
$$
\frac{\Gamma ,x:K\vdash k\colon K'}{\Gamma \vdash [x:K]k\colon
(x:K)K'}\: \: \: \: \: \: \frac{\Gamma \vdash K_{1}=K_{2}\: \: \:
\Gamma ,x:K_{1}\vdash k_{1}=k_{2}\colon K}{\Gamma \vdash
[x:K_{1}]k_{1}=[x:K_{2}]k_{2}\colon (x:K_{1})K}$$
$$
\frac{\Gamma \vdash f\colon (x:K)K'\: \: \Gamma \vdash k\colon
K}{\Gamma \vdash f(k)\colon [k/x]K'}\: \: \: \frac{\Gamma \vdash
f=f'\colon (x:K)K'\: \: \Gamma \vdash k_{1}=k_{2}\colon K}{\Gamma
\vdash f(k_{1})=f'(k_{2})\colon [k_{1}/x]K'}$$
$$
\: \frac{\Gamma ,x:K\vdash k'\colon K'\: \: \Gamma \vdash k\colon
K}{\Gamma \vdash ([x:K]k')(k)=[k/x]k'\colon [k/x]K'}\: \: \: \:
\frac{\Gamma \vdash f\colon (x:K)K'\: \: x\notin FV(f)}{\Gamma
\vdash [x:K]f(x)=f\colon (x:K)K'}$$

\selfc{
\begin{figure}[top]
\framebox[5.8in][l]{
\begin{minipage}{\linewidth}
%\ \\
\ \ \ \emph{Contexts and assumptions}
$$
% empty context I change it from <> to ()  --
\frac{}{ () \: \: valid}\: \: \: \frac{\Gamma \vdash K\:
\: kind\: \: \: x\notin FV(\Gamma )}{\Gamma ,x:K\: \: valid}\: \: \:
\frac{\Gamma ,x:K,\Gamma '\: \: valid}{\Gamma ,x:K,\Gamma '\vdash
x\colon K}$$
\ \ \ \emph{General equality rules}
$$
\frac{\Gamma \vdash K\: \: kind}{\Gamma \vdash K=K}\: \: \:
\frac{\Gamma \vdash K=K'}{\Gamma \vdash K'=K}\: \: \: \frac{\Gamma
\vdash K=K'\: \: \Gamma \vdash K'=K''}{\Gamma \vdash K=K''}$$
$$
\frac{\Gamma \vdash k\colon K}{\Gamma \vdash k=k\colon K}\: \: \:
\frac{\Gamma \vdash k=k'\colon K}{\Gamma \vdash k'=k\colon K}\: \:
\: \frac{\Gamma \vdash k=k'\colon K\: \: \Gamma \vdash k'=k''\colon
K}{\Gamma \vdash k=k''\colon K}$$
\ \ \ \emph{Equality typing rules}
$$
\frac{\Gamma \vdash k\colon K\: \: \Gamma \vdash K=K'}{\Gamma \vdash
k\colon K'}\: \: \: \frac{\Gamma \vdash k=k'\colon K\: \: \Gamma
\vdash K=K'}{\Gamma \vdash k=k'\colon K'}$$
\ \ \ \emph{Substitution rules}
$$
\frac{\Gamma ,x:K,\Gamma '\: \: valid\: \: \Gamma \vdash k\colon
K}{\Gamma ,[k/x]\Gamma '\: \: valid}$$
$$
\frac{\Gamma ,x:K,\Gamma '\vdash K'\: \: kind\: \: \Gamma \vdash
k\colon K}{\Gamma ,[k/x]\Gamma '\vdash [k/x]K'\: \: kind}\: \: \:
\frac{\Gamma ,x:K,\Gamma '\vdash K'\: \: kind\: \: \Gamma \vdash
k=k'\colon K}{\Gamma ,[k/x]\Gamma '\vdash [k/x]K'=[k'/x]K'}$$
$$
\frac{\Gamma ,x:K,\Gamma '\vdash k'\colon K'\: \: \Gamma \vdash
k\colon K}{\Gamma ,[k/x]\Gamma '\vdash [k/x]k'\colon [k/x]K'}\: \:
\: \frac{\Gamma ,x:K,\Gamma '\vdash k'\colon K'\: \: \Gamma \vdash
k_{1}=k_{2}\colon K}{\Gamma ,[k_{1}/x]\Gamma '\vdash
[k_{1}/x]k'=[k_{2}/x]k'\colon [k_{1}/x]K'}$$
$$
\frac{\Gamma ,x:K,\Gamma '\vdash K'=K''\: \: \Gamma \vdash k\colon
K}{\Gamma ,[k/x]\Gamma '\vdash [k/x]K'=[k/x]K''}\: \: \:
\frac{\Gamma ,x:K,\Gamma '\vdash k'=k''\colon K'\: \: \Gamma \vdash
k\colon K}{\Gamma ,[k/x]\Gamma '\vdash [k/x]k'=[k/x]k''\colon
[k/x]K'}$$
\ \ \ \emph{The kind Type}
$$
\frac{\Gamma \: \: valid}{\Gamma \vdash Type\: \: kind}\: \: \:
\frac{\Gamma \vdash A\colon Type}{\Gamma \vdash El(A)\: \: kind}\:
\: \: \frac{\Gamma \vdash A=B\colon Type}{\Gamma \vdash
El(A)=El(B)}$$
\ \ \ \emph{Dependent product kinds}
$$
\frac{\Gamma \vdash K\: \: kind\: \: \Gamma ,x:K\vdash K'\: \:
kind}{\Gamma \vdash (x:K)K'\: \: kind}\: \: \: \frac{\Gamma \vdash
K_{1}=K_{2}\: \: \Gamma ,x:K_{1}\vdash K_{1}'=K_{2}'}{\Gamma \vdash
(x:K_{1})K_{1}'=(x:K_{2})K_{2}'}$$
$$
\frac{\Gamma ,x:K\vdash k\colon K'}{\Gamma \vdash [x:K]k\colon
(x:K)K'}\: \: \: \: \: \: \frac{\Gamma \vdash K_{1}=K_{2}\: \: \:
\Gamma ,x:K_{1}\vdash k_{1}=k_{2}\colon K}{\Gamma \vdash
[x:K_{1}]k_{1}=[x:K_{2}]k_{2}\colon (x:K_{1})K}$$
$$
\frac{\Gamma \vdash f\colon (x:K)K'\: \: \Gamma \vdash k\colon
K}{\Gamma \vdash f(k)\colon [k/x]K'}\: \: \: \frac{\Gamma \vdash
f=f'\colon (x:K)K'\: \: \Gamma \vdash k_{1}=k_{2}\colon K}{\Gamma
\vdash f(k_{1})=f'(k_{2})\colon [k_{1}/x]K'}$$
$$
\: \frac{\Gamma ,x:K\vdash k'\colon K'\: \: \Gamma \vdash k\colon
K}{\Gamma \vdash ([x:K]k')(k)=[k/x]k'\colon [k/x]K'}\: \: \: \:
\frac{\Gamma \vdash f\colon (x:K)K'\: \: x\notin FV(f)}{\Gamma
\vdash [x:K]f(x)=f\colon (x:K)K'}$$
%\\
\end{minipage}
}%\framebox
\caption{Inference Rules of Logical Framework} \label{LF-rules}
\end{figure}
}%selfc

%\newpage
\section{Inference Rules of Typed Operational Semantics for LF}
\label{app:LF-TOSrules}

The inference rules of the TOS for LF are given below.  (See \cite{healf:TLCA99YY,healf:thesis} for further details.)

\ \\

\ \ \ \emph{Contexts}
$$
\frac{}{\models () \rightarrow ()} \ EMP \: \: \: \: \: \:
\frac{\models \Gamma \rightarrow \Delta \: \: \: \Gamma \models A \rightarrow B \:
\: \: x \notin dom\{\Gamma\}}{\models \Gamma, x:A \rightarrow \Delta, x:B} \ WEAK
$$

\ \ \ \emph{Kinds}
$$
\frac{\Gamma \models ok}{\Gamma \models Type \rightarrow Type} \ TYPE \: \: \: \: \: \:
\frac{\Gamma \models M \rightarrow N \rightarrow P \colon Type}{\Gamma \models El(M)
\rightarrow El(P)} \ EL
$$
\\
$$
\frac{\Gamma \models A_1 \rightarrow B_1 \: \: \: \Gamma, x: A_1 \models A_2
\rightarrow B_2}{\Gamma \models (x:A_1)A_2 \rightarrow (x:B_1).B_2} \ PI
$$

\ \ \ \emph{Terms}
$$
\frac{\Gamma_0, x:A, \Gamma_1 \models A \rightarrow B}{\Gamma_0, x:A, \Gamma_1
\models x \rightarrow x \rightarrow x \colon B} \ VAR
$$
\\
$$
\frac{\Gamma \models A_1 \rightarrow B_1 \: \: \: \Gamma, x:A_1 \models M_0
\rightarrow_{n} P_0 \colon B_2 \: \: \: [x:B_1]P_0 \ not \ \eta -redex}{\Gamma
\models [x:A_1]M_0 \rightarrow [x:A_1]M_0 \rightarrow [x:B_1]P_0 \colon (x:B_1)B_2}
\ LAM
$$
\\
$$
\frac{\Gamma \models A_1 \rightarrow B_1 \: \: \: \Gamma, x:A_1 \models M_0
\rightarrow_{n} P(x)\colon B_2 \: \: \: \Gamma \models P \rightarrow P \rightarrow P
\colon (x:B_1)B_2}{\Gamma \models [x:A_1]M_0 \rightarrow [x:A_1]M_0 \rightarrow P
\colon (x:B_1)B_2} \ ETA
$$
\\
$$
\frac{\begin{array}{c} \Gamma \models M_1 \rightarrow N_1 \rightarrow P_1 \colon
(x:B_1)B_2 \: \: \: \Gamma \models M_2 \rightarrow N_2 \rightarrow P_2 \colon B_1 \:
\: \: \\ \Gamma \models [M_2/x]B_2 \rightarrow C \: \: \: N_1 \ not \ abstraction
\end{array}}{\Gamma \models M_1(M_2) \rightarrow N_1(M_2) \rightarrow P_1(P_2)
\colon C} \ BASE
$$
\\
$$
\frac{\begin{array}{c} \Gamma \models M_1 \rightarrow_{w} [x:A_1]N_0 \colon
(x:B_1)B_2 \: \: \: \Gamma \models M_2 \colon B_1 \: \: \: \\ \Gamma \models
[M_2/x]N_0 \rightarrow P \rightarrow Q \colon C \: \: \: \Gamma \models [M_2/x]B_2
\rightarrow C \end{array}}{\Gamma \models M_1(M_2) \rightarrow P \rightarrow Q
\colon C} \ BETA
$$

%%% This is the changed Figure 6 (May 17th) with lambda PI notation %%%
\selfc{
\begin{figure}[here] % here
\framebox[6.2in][l]{
\begin{minipage}{\linewidth}
\ \\

\ \ \ \emph{Contexts}
%\Frule{}{\models () \rightarrow ()}{EMP}{}
%\Frule{\models \Gamma \rightarrow \Delta \: \: \: \Gamma \models A \rightarrow B \: \: \: x \notin dom\{\Gamma\}}{\models \Gamma, x:A \rightarrow \Delta, x:B}{WEAK}{}
$$
\frac{}{\models () \rightarrow ()} \ EMP \: \: \: \: \: \:
\frac{\models \Gamma \rightarrow \Delta \: \: \: \Gamma \models A \rightarrow B \: \: \: x \notin dom\{\Gamma\}}{\models \Gamma, x:A \rightarrow \Delta, x:B} \ WEAK
$$

\ \ \ \emph{Kinds}
$$
\frac{\Gamma \models ok}{\Gamma \models Type \rightarrow Type} \ TYPE \: \: \: \: \: \:
\frac{\Gamma \models M \rightarrow N \rightarrow P \colon Type}{\Gamma \models El(M) \rightarrow El(P)} \ EL
$$
\\
$$
\frac{\Gamma \models A_1 \rightarrow B_1 \: \: \: \Gamma, x: A_1 \models A_2 \rightarrow B_2}{\Gamma \models \Pi x:A_1.A_2 \rightarrow \Pi x:B_1.B_2} \ PI
$$

\ \ \ \emph{Terms}
$$
\frac{\Gamma_0, x:A, \Gamma_1 \models A \rightarrow B}{\Gamma_0, x:A, \Gamma_1 \models x \rightarrow x \rightarrow x \colon B} \ VAR
$$
\\
$$
\frac{\Gamma \models A_1 \rightarrow B_1 \: \: \: \Gamma, x:A_1 \models M_0 \rightarrow_{n} P_0 \colon B_2 \: \: \: \lambda x:B_1.P_0 \ not \ \eta -redex}{\Gamma \models \lambda x:A_1.M_0 \rightarrow \lambda x:A_1.M_0 \rightarrow \lambda x:B_1.P_0 \colon \Pi x:B_1.B_2} \ LAM
$$
\\
$$
\frac{\Gamma \models A_1 \rightarrow B_1 \: \: \: \Gamma, x:A_1 \models M_0 \rightarrow_{n} P(x)\colon B_2 \: \: \: \Gamma \models P \rightarrow P \rightarrow P \colon \Pi x:B_1.B_2}{\Gamma \models \lambda x:A_1. M_0 \rightarrow \lambda x:A_1.M_0 \rightarrow P \colon \Pi x:B_1.B_2} \ ETA
$$
\\
$$
\frac{\begin{array}{c} \Gamma \models M_1 \rightarrow N_1 \rightarrow P_1 \colon \Pi x:B_1.B_2 \: \: \: \Gamma \models M_2 \rightarrow N_2 \rightarrow P_2 \colon B_1 \: \: \: \\ \Gamma \models [M_2/x]B_2 \rightarrow C \: \: \: N_1 \ not \ abstraction \end{array}}{\Gamma \models M_1(M_2) \rightarrow N_1(M_2) \rightarrow P_1(P_2) \colon C} \ BASE
$$
\\
$$
\frac{\begin{array}{c} \Gamma \models M_1 \rightarrow_{w} \lambda x:A_1.N_0 \colon \Pi x:B_1.B_2 \: \: \: \Gamma \models M_2 \colon B_1 \: \: \: \\ \Gamma \models [M_2/x]N_0 \rightarrow P \rightarrow Q \colon C \: \: \: \Gamma \models [M_2/x]B_2 \rightarrow C \end{array}}{\Gamma \models M_1(M_2) \rightarrow P \rightarrow Q \colon C} \ BETA
$$
\\
\end{minipage}
}%\framebox
\caption{Inference Rules of Typed Operational Semantics for LF} \label{TOS-LF-rules}
\end{figure}

}%selfc

\end{document}